\documentclass[journal]{IEEEtran}

\usepackage{cite}
\usepackage[pdftex]{graphicx}

\DeclareGraphicsExtensions{.pdf,.jpeg, .jpg, .png, .eps}
\graphicspath{{./figures/}}
\usepackage{stfloats}
\usepackage{url}
\usepackage{multirow}
\usepackage[table, dvipsnames]{xcolor}
\usepackage{tabularx}
\usepackage{hyperref}
\usepackage{booktabs}
\usepackage[caption=false,font=footnotesize]{subfig}
\usepackage[nameinlink]{cleveref}
\Crefname{equation}{Eq.}{Eqs.}
\Crefname{figure}{Fig.}{Figs.}
\Crefname{tabular}{Tab.}{Tabs.}
\usepackage{enumitem}

\usepackage{balance}

\usepackage{tikz}
\usetikzlibrary{tikzmark, calc}
\usepackage{dcolumn, array, booktabs}
\usepackage{boldline}
\usepackage[normalem]{ulem}

\usepackage{scalerel}
\usepackage{tikz}
\usetikzlibrary{svg.path}

\definecolor{orcidlogocol}{HTML}{A6CE39}
\tikzset{
  orcidlogo/.pic={
    \fill[orcidlogocol] svg{M256,128c0,70.7-57.3,128-128,128C57.3,256,0,198.7,0,128C0,57.3,57.3,0,128,0C198.7,0,256,57.3,256,128z};
    \fill[white] svg{M86.3,186.2H70.9V79.1h15.4v48.4V186.2z}
                 svg{M108.9,79.1h41.6c39.6,0,57,28.3,57,53.6c0,27.5-21.5,53.6-56.8,53.6h-41.8V79.1z M124.3,172.4h24.5c34.9,0,42.9-26.5,42.9-39.7c0-21.5-13.7-39.7-43.7-39.7h-23.7V172.4z}
                 svg{M88.7,56.8c0,5.5-4.5,10.1-10.1,10.1c-5.6,0-10.1-4.6-10.1-10.1c0-5.6,4.5-10.1,10.1-10.1C84.2,46.7,88.7,51.3,88.7,56.8z};
  }
}

\newcommand\orcidicon[1]{\href{https://orcid.org/#1}{\mbox{\scalerel*{
\begin{tikzpicture}[yscale=-1,transform shape]
\pic{orcidlogo};
\end{tikzpicture}
}{|}}}}

\usepackage{hyperref} 

\tikzset{
    double color fill/.code 2 args={
        \pgfdeclareverticalshading[%
            tikz@axis@top,tikz@axis@middle,tikz@axis@bottom%
        ]{diagonalfill}{100bp}{%
            color(0bp)=(tikz@axis@bottom);
            color(50bp)=(tikz@axis@bottom);
            color(50bp)=(tikz@axis@middle);
            color(50bp)=(tikz@axis@top);
            color(100bp)=(tikz@axis@top)
        }
        \tikzset{shade, left color=#1, right color=#2, shading=diagonalfill}
    }
}
\newcommand\BGcell[4][0pt]{%
  \begin{tikzpicture}[overlay,remember picture]%
    \path[#4] ( $ (pic cs:#2) + (-.5\tabcolsep,1.9ex) $ ) rectangle ( $ (pic cs:#3) + (\tabcolsep,-#1*\baselineskip-.8ex) $ );
  \end{tikzpicture}%
}%
\newcounter{BGnum}
\newcommand\cellBG[4]{
    \multicolumn{#1}{
        !{\BGcell{startBG\arabic{BGnum}}{endBG\arabic{BGnum}}{%
                #2}
            \tikzmark{startBG\arabic{BGnum}}}
            #3
        !{\tikzmark{endBG\arabic{BGnum}}}}
        {#4} 
      \addtocounter{BGnum}{1}
}

\newcolumntype{.}{D{.}{.}{-1}}

\tikzset{%
    diagonal fill/.style 2 args={%
        double color fill={#1}{#2},
        shading angle=45,
        opacity=0.8}
   }

\definecolor{IS_Blue}{HTML}{A9C4EB}
\definecolor{WS_Green}{HTML}{B9E0A5}
\definecolor{OS_Red}{HTML}{F19C99}
\definecolor{BG_Purple}{HTML}{C3ABD0}
\definecolor{Grey}{HTML}{CCCCCC}

\begin{document}

\title{Taxonomy and Benchmarking of \\Precision-Scalable MAC Arrays under \\Enhanced DNN Dataflow Representation}

\author{
    \IEEEauthorblockN{Ehab~M.~Ibrahim$^\dagger$~\orcidicon{0000-0003-1250-0160}, Linyan Mei$^\dagger$~\orcidicon{0000-0001-8649-3923},~\IEEEmembership{Student~Member,~IEEE}, and~Marian~Verhelst~\orcidicon{0000-0003-3495-9263},~\IEEEmembership{Senior~Member,~IEEE}}\\\vspace{0.3cm}\footnotesize{$^\dagger$These~authors~contributed~equally~to~this~work.\vspace{-0.3cm}}
\vspace{-0.1em}
\thanks{
The authors are with MICAS Laboratories, Electrical Engineering Department (ESAT), KU Leuven, Belgium. (Emails: ehab.ibrahim@magics.tech; linyan.mei@kuleuven.be; marian.verhelst@kuleuven.be)

This work was supported in part by the Flemish Government (AI Research Program), in part by the Fund For Scientific Research Flanders (FWO-Vlaanderen) under Grant G006718N, and in part by the EU under Grant ERC-2016-STG-715037.

This work was done when Ehab M. Ibrahim was a research assistant in ESAT-MICAS, KU Leuven from 2020 to 2021.

}
}

\maketitle

\begin{abstract}

Reduced-precision and variable-precision multiply-accumulate (MAC) operations provide opportunities to significantly improve energy efficiency and throughput of DNN accelerators with no/limited algorithmic performance loss, paving a way towards deploying AI applications on resource-constraint edge devices. Accordingly, various precision-scalable MAC array (PSMA) architectures were proposed recently. However, it is difficult to make a fair comparison between those alternatives, as each proposed PSMA is demonstrated in different systems and technologies. This work aims to provide a clear view of the design space of PSMA and offer insights for selecting the optimal architectures based on designers' needs. First, we introduce a precision-enhanced for-loop representation for DNN dataflows. Next, we use this new representation towards a comprehensive PSMA taxonomy, capable of systematically covering most prominent state-of-the-art PSMAs, as well as uncovering new PSMA architectures. Following that, we build a highly parameterized PSMA template that can be design-time configured into a huge subset of the design space spanned by the taxonomy. This allows to fairly and thoroughly benchmark 72 different PSMA architectures. We perform such studies in 28nm technology targeting run-time precision scalability from 8 to 2 bits, operating at 200 MHz and 1 GHz. Analyzing resulting energy and area breakdowns reveals key design guidelines for PSMA architecures.

\end{abstract}


\bstctlcite{IEEEexample:BSTcontrol}

\section{Introduction}\label{sec:Intro}
  Deep learning inference on edge devices is quickly gaining traction. However, the main hurdle holding back the widespread deployment of embedded inference is the limited available energy budget at the edge. Researchers are hence focusing on techniques improving energy efficiency of embedded deep neural network (DNN) inference. A popular approach to achieve high energy efficiency is reducing the precision of the DNN operands (weight, input and output activation), more commonly known as network quantization. Due to its computational nature, DNNs are rather resilient to approximation errors \cite{chippa2013analysis,Soft_Error}, and accordingly, can reduce parameter precision without much penalty in inference accuracy \cite{Moons2016Quantized,hubara2017quantized,cavigelli2016origami}. 
  
  There are two approaches to DNN quantization: 1.) Post-training quantization, which trains the network on full-precision, then quantizes the pre-trained parameters (with/without post fine-tuning) \cite{judd2018proteus}; 2.) Quantization-aware training, which  directly trains the network with lower precisions \cite{courbariaux2014training}. These approaches progressively enabled DNNs to first be quantized to 16-bit fixed point \cite{gupta2015deep}, 8-bit fixed point \cite{banner2018scalable}, and all the way down to binary precision \cite{courbariaux2015binaryconnect}. The best precision of DNN parameters, however, varies across different NN models, and even across different layers within one model \cite{judd2018proteus, Zhang2020Precision}. This leads to the trend of transferring from the traditional fixed-precision MAC array to run-time precision-scalable MAC arrays (PSMAs) in DNN accelerators, which can at run-time tune to the desired precision mode so as to effectively translate those low precision opportunities into energy and performance gains.

  Many precision-scalable accelerators were proposed in recent years, some of which support weight-only (1D) precision scalability \cite{shin2017dnpu, lee2018unpu, TCAS_I}, or input/weight (2D) precision scalability \cite{sharma2018bitfusion, ryu2019bitblade, sharify2018loom, ghodrati2020bit, BitSystolic}. As all these precision scalable MACs have been demonstrated with different RTL coding styles, in different silicon technologies, in different MAC array configurations, and with different DNN workloads, it is hard to relatively compare them in terms of their drawbacks and merits.
  
  A comprehensive survey and comparison of precision-scalable MAC units was presented in \cite{camus2019review} at the MAC level. However, this work only focused on a single MAC, and did not study the implications of using such MACs in an array. This study hence omitted two important efficiency factors when integrating such MACs in a larger MAC array: 1.) Depending on the precision-scalable MAC architecture, the scalability overhead can, or cannot, be amortized across all MACs of the MAC array, strongly impacting resulting energy and area overheads per operation; 2.) The precision-scalable MAC architecture has an impact on which spatial unrolling dimensions can be exploited across the MAC array in low precision modes, again resulting in array-level energy efficiency implications.
  
  To overcome these shortcomings and enable a fair comparison of precision-scalable architectures at the MAC array level, this paper makes the following contributions:
  
  \begin{itemize}
      \item Extension of the traditional 7 for-loop representation of convolutional neural networks (CNNs) layer with 2 additional precision-enhanced bit-group loops. Discussion of the implications of the two added bit-group loops on the underlying hardware and their spatial/temporal unrolling options (\Cref{sec:CNN_loops}).
      \item Introduction of a new PSMA taxonomy, based on the newly proposed 9-loop CNN representation. Mapping of a wide range of start-of-the-art (SotA) PSMA architectures to the new taxonomy is conducted, and new topologies are identified (\Cref{sec:taxonomy}).
      \item Design of a uniform and highly parameterized PSMA template, which can be design-time configured into a huge subset of the design space spanned by the taxonomy. (\Cref{sec:uniform_mac}).
      \item Benchmarking this design space in terms of area and energy, resulting in PSMA design insights. (\Cref{sec:comparative_study}).
  \end{itemize}

Source code and supplementary materials are available at: \href{https://github.com/KULeuven-MICAS/PSMA_benchmark}{\uline{https://github.com/KULeuven-MICAS/PSMA-benchmark}}.

\section{Precision-Enhanced CNN Loop Representation} \label{sec:CNN_loops}
  This section introduces a precision-driven extension to the traditional 7 CNN for-loop format that has been widely used to represent the dataflow of different accelerators. First, the motivation behind this extension is highlighted, after which the resulting loops are categorized based on their effects on the underlying hardware. 
  
  \subsection{Motivation}
  To fully understand the motivation of extending the traditional 7 CNN loop representation, the concept of bit-groups (BGs) is first explained. Precision-scalable accelerators typically compose the underlying MAC units out of reduced precision sub-multipliers \cite{shun2007booth}. The intermediate partial products are then either shifted and added together to achieve the full precision result, or treated as separate results when the MAC operates in its low precision mode. 
  
  \begin{figure}[!tb]
  \centering
  \includegraphics[width=0.48\textwidth]{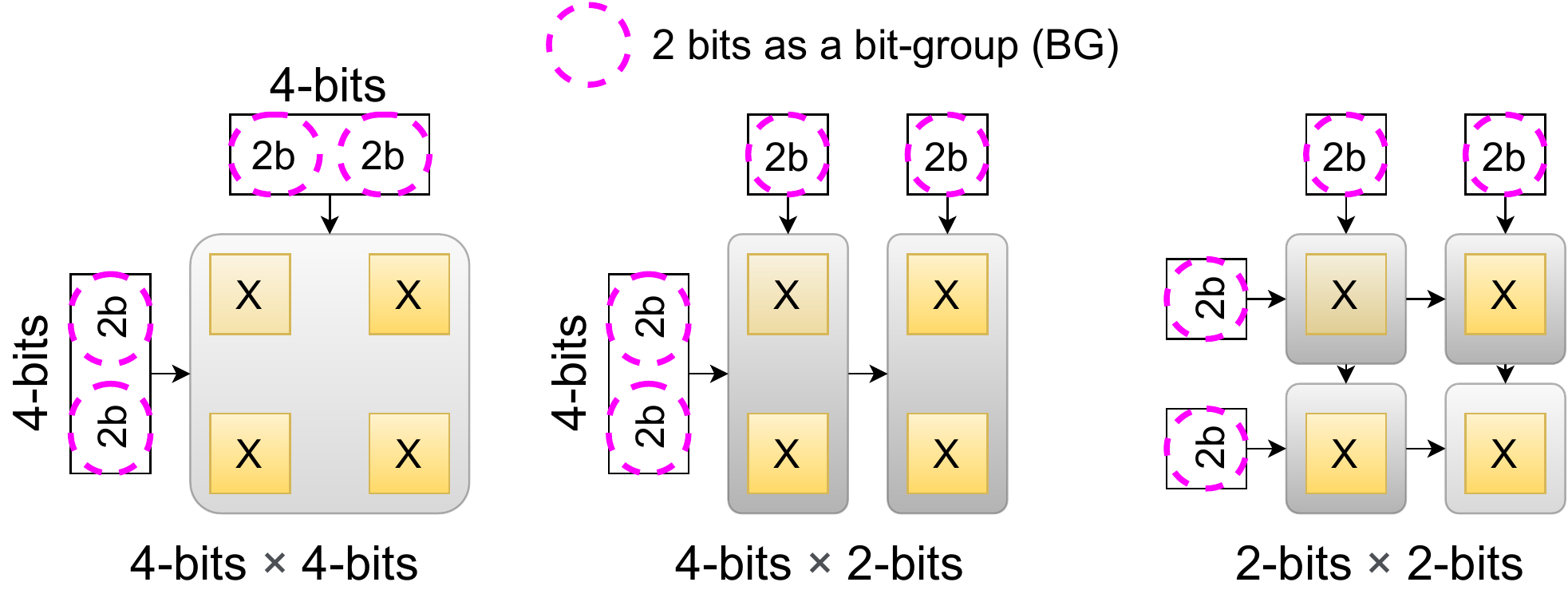}
  \caption{A simple 4-bit precision scalable multiplier.}
  \label{fig:4-bit_mult}
  \end{figure}

  To illustrate this, a simple 4-bit precision-scalable multiplier is shown in \Cref{fig:4-bit_mult}, which has a similar architecture to the BitBricks found in BitFusion \cite{sharma2018bitfusion}. In the shown example, the 4$\times$4-bit multiplier consists of 4 sub-units, each being a 2$\times$2-bit multiplier. In full 4-bit precision, both inputs are divided into 2 Bit Groups (BGs) of 2-bits each, which together form the full 4-bit input resolution. These BGs are multiplied in the different sub-units of the multiplier, and get shifted and aggregated to obtain the final 4$\times$4-bit multiplication result. In the low precision 2-bit mode, 4 inputs of 2-bits are each used independently, hence each consisting of only a single BG. In this case, each sub-unit acts as a standalone multiplier, producing an independent result. In a sense, the multiplier is hence considered as a mini-array that performs spatial unrolling at the BG level at full precision, and enables additional spatial unrolling capabilities at low precision.

\begin{figure}[!tb]
    \centering
    \includegraphics[width=0.48\textwidth]{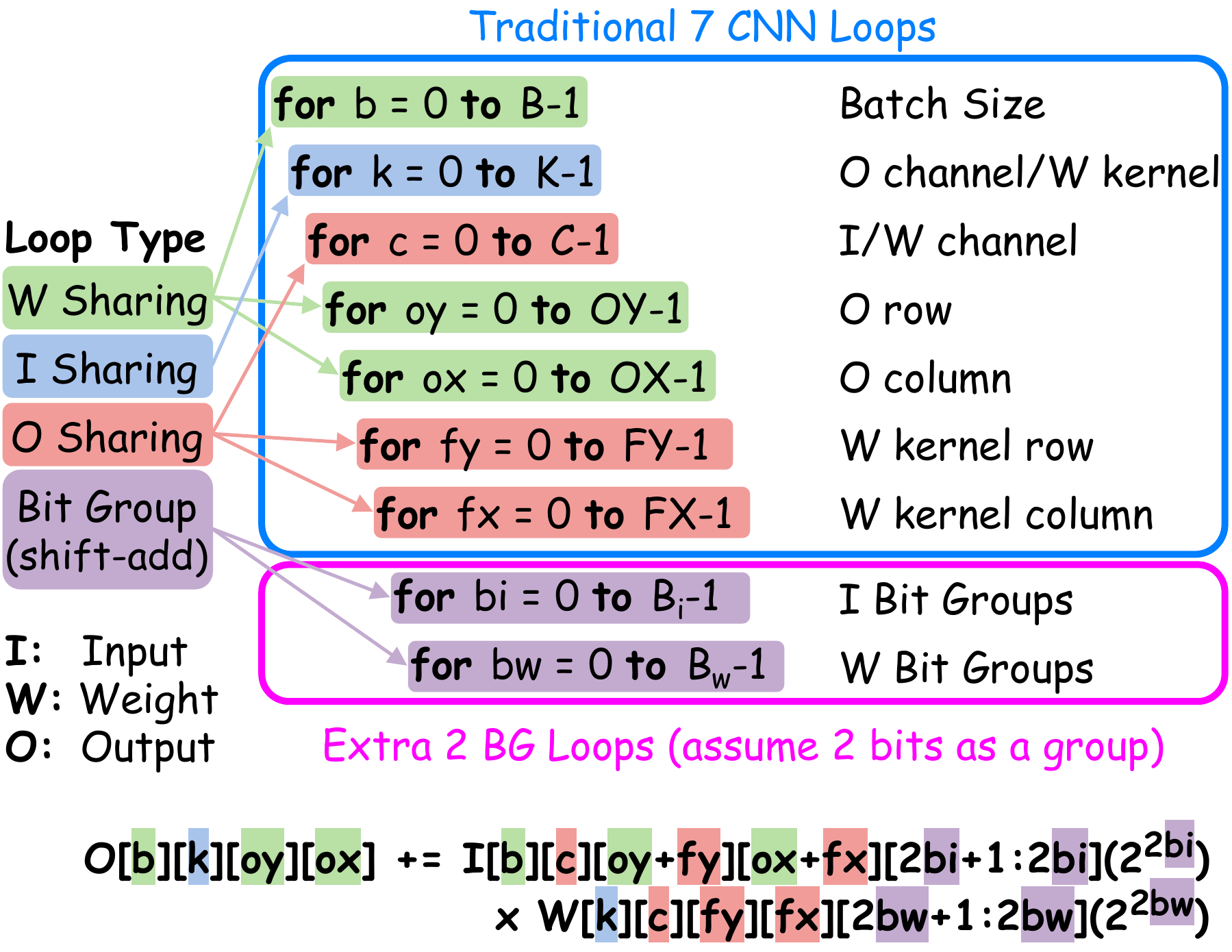}
    \caption{Introducing 2 extra bit-group (BG) for-loops.}
    \label{fig:CNN_loops}
  \end{figure}

  This concept can now be introduced in the CNN dataflow representation to unify across all precision scalability techniques. The CNN accelerator dataflow, which refers to the spatial unrolling (how CNN loops are spatially unrolled/mapped onto the 2D MAC array) and temporal mapping (how CNN loops are temporally scheduled/mapped upon the accelerator) \cite{mei2021zigzag}, is traditionally characterized as 7 nested for-loops. To include the impact of precision scalability into this representation, 2 additional BG for-loops are added to the traditional 7 CNN for-loops, as shown in \Cref{fig:CNN_loops}. In the previous example of \Cref{fig:4-bit_mult}, at high precisions, the 2 BG loops are spatially unrolled inside the MAC unit. In this mode, the MAC unit is hence unable to unroll any other loop spatially. While at lower precisions, the number of BG loops of the workload shrinks, making room for other for-loops to be spatially unrolled inside of the MAC unit.
  
  Treating the BG loops similar to any other CNN loop enables the flexibility of unrolling them spatially and also temporally, as done in bit-serial (BS) architectures. This allows us to uniformly characterize the behaviour of a wide variety of PSMAs. 
  Note that besides convolutional layers, other DNN layers like fully-connected, depthwise and pointwise can also be extended following the same method.
  
  \subsection{Implications on the MAC Array Hardware}
  In addition to BG loops, it is essential to understand how each loop can be mapped to a MAC array. As shown in \Cref{fig:CNN_loops}, the 7 traditional CNN loops are categorized based on input, weight, or output sharing.
  For example, `K' (output channel dimension) is an input sharing loop dimension in CNN computation because when looping through `K' (spatially or temporally), the same input elements are required and thus shareable. The same concept applies for weight and output sharing loops.
  
  When input or weight sharing loops are spatially unrolled across a dimension of a MAC array, the inputs/weights can be broadcast to all the MAC units along that dimension of the MAC array. On the other hand, when an output-sharing loop is spatially unrolled, the outputs of each MAC unit are added together along that dimension. Additionally, spatial or temporal unrolling of BG loops has hardware implications, as it would require the partial results to go through a shift \& add tree or get accumulated in a shift-add register, respectively.
  
\section{Precision-Scalable MAC Array Taxonomy} \label{sec:taxonomy}
  Making use of the new loop representation, a taxonomy will be introduced, which can be used to map all SotA precision-scalability techniques. This new taxonomy will serve as a definition for the full design space of PSMAs, and will later be explored and benchmarked extensively in \Cref{sec:comparative_study}.

  \subsection{Highly Parameterized PSMA Template} \label{subsec:approach}

  \begin{figure}[tb]
    \centering
    \includegraphics[width=0.48\textwidth]{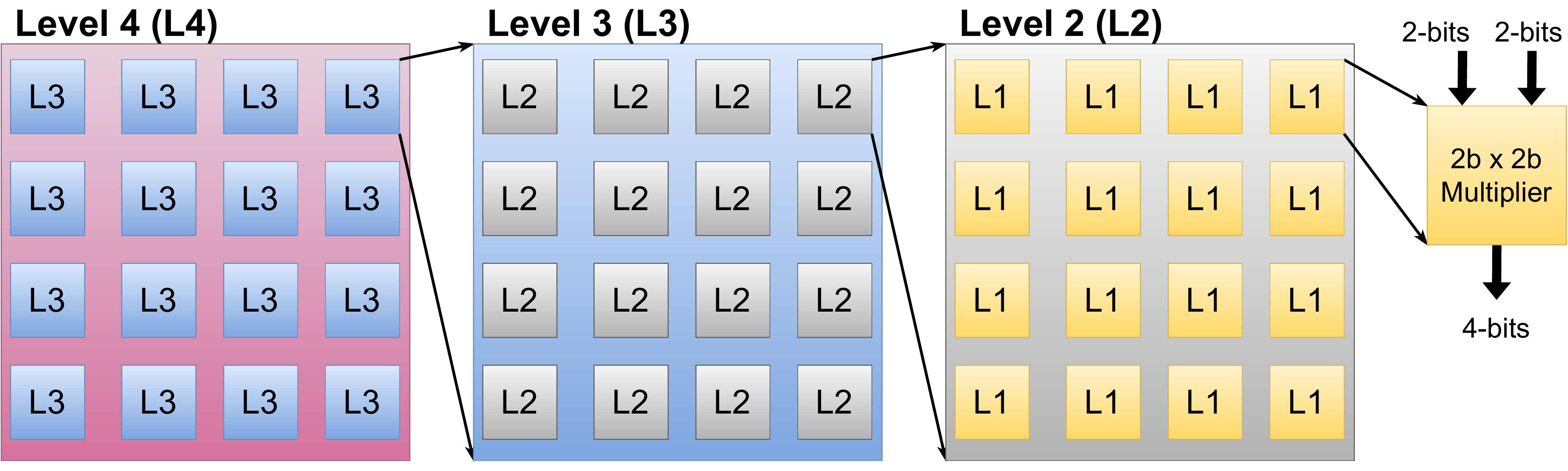}
    \caption{L4: 4$\times$4 L3 units; L3: 4$\times$4 L2 units; L2: 4$\times$4 L1 units.}
    \label{fig:layered_MAC}
  \end{figure}
  
  \begin{figure}[tb]
    \centering
    \includegraphics[width=0.48\textwidth]{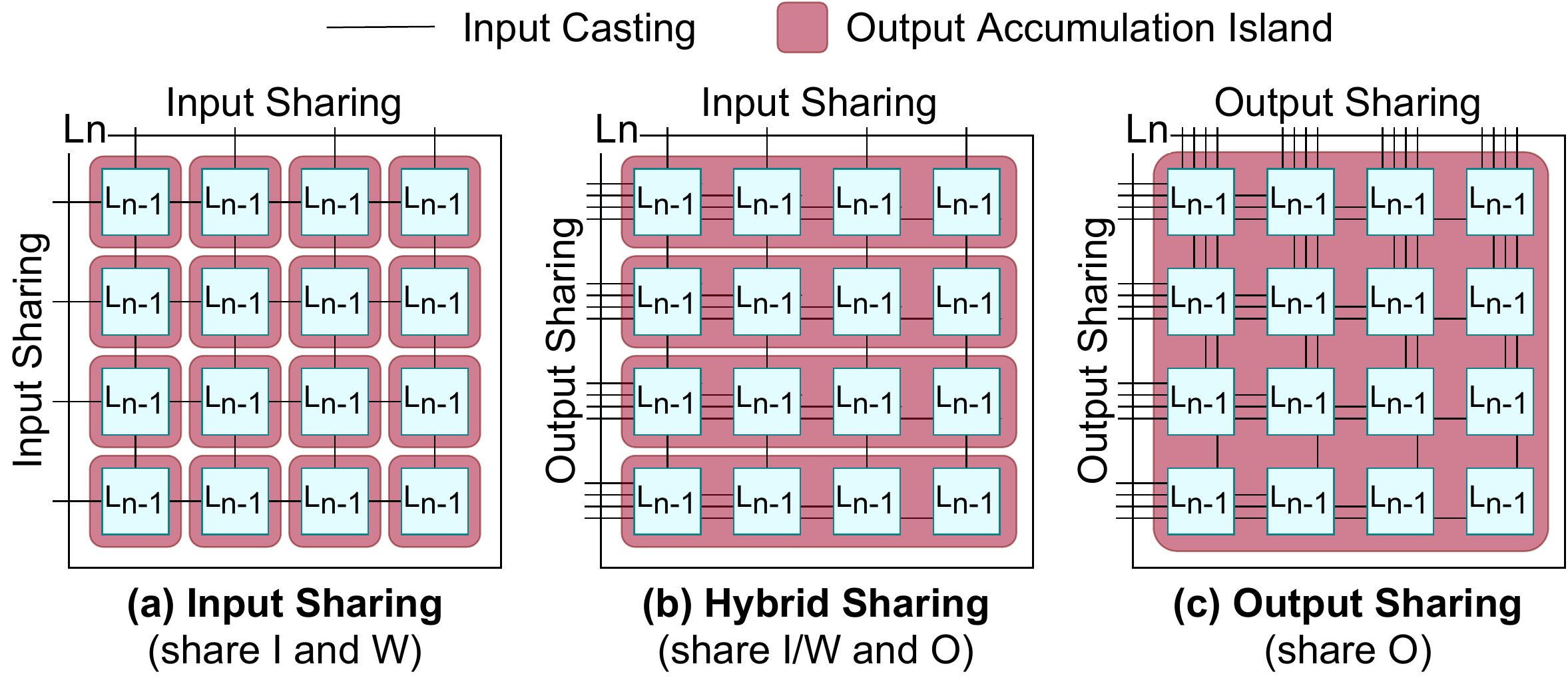}
    \caption{Input, Hybrid, and Output Sharing at Level n (Ln). Results from one accumulation island are added together spatially to form one output.}
    \label{fig:acc_islands}
    \vspace{-1em}
  \end{figure}
  
  A uniform and highly parameterized MAC-array architecture template, which is compatible with the precision-scalability concepts introduced in previous sections, is built up from basic 2-bit multiplier building blocks (denoted here as ``Level 1 (L1)'' units). These basic L1 units are subsequently combined in different hierarchical levels to form the complete MAC array. We choose to combine 4$\times$4 units in each step, hence resulting in 16 L1 units forming one L2 unit, 16 L2 units forming one L3 unit, 16 L3 units forming one L4 unit. An illustration of the proposed template is shown in \Cref{fig:layered_MAC}.
  
  Flexibility is the main motive for going with such a hierarchical architecture. With such a template, one can stop at L3 for a smaller MAC array size, go for L4 for a standard array, or even extend to higher levels for a larger array. Additionally, having a small L1 unit allows us to have precision-scalable MAC units that can go down to 2-bit precision, which is the lowest precision supported in most precision-scalable accelerators \cite{sharma2018bitfusion, ryu2019bitblade, ghodrati2020bit, moons2017envision}. Furthermore, this multi-level architecture will later give a clear perspective of where each CNN loop gets unrolled to.
  
  \subsection{Spatial Unrolling}
  
    The CNN loop categorization of previous section (Weight sharing (WS), Input sharing (IS), Output sharing (OS), and BG loops), can now be linked to the hierarchical array template to assess the hardware consequences. We define that every level $L_n$ of the MAC array template can support a specific spatial unrolling along its horizontal dimension, and one along its vertical dimension (which can be identical to or different from the horizontal one). Specifically, if IS/WS loops are spatially unrolled along a dimension of a certain level $L_{n}$, it means that the inputs/weights are broadcast across the different $L_{n-1}$ units of this dimension. This also implies that the outputs of the $L_{n-1}$ units in this dimension can not be added together, and need separate accumulators. 
    
    On the other hand, if an OS loop is spatially unrolled along a dimension of $L_n$, the partial sums of the $L_{n-1}$ units across that dimension are added together with an adder tree. This then implies that the inputs and weights of all these $L_{n-1}$ units on that dimension are unique.
    
    The resulting 3 unrolling possibilities are illustrated in \Cref{fig:acc_islands}. One level in the MAC array template hierarchy will either: a.) share inputs along both dimensions (inputs and weights), resulting in 16 independent outputs; b.) share inputs along one dimension (inputs or weights) and outputs along another one, resulting in 4 aggregated outputs; c.) share outputs along both dimensions, resulting in 1 joint sum.
  
  \subsection{Bit-Group Unrolling}
  Besides WS/IS/OS loops, the BG loops can also be spatially unrolled on the MAC array, and dictate where the shift \& add tree needs to be inserted. In the proposed taxonomy, BG loop unrolling can either occur spatially at one of the two lowest levels of the MAC array (so either at the L2 or at the L3 level; note that L1 is not an array level), or can occur temporally, a.k.a. bit-serial processing (BS). For cases in which BG loops are unrolled temporally, internal registers are required in order to hold intermediate results towards performing shift-add operation temporally. BS designs can be further sub-categorized based on the level in the PSMA at which the internal registers are located.
  
  Note that BG loops are only unrolled at precisions higher than the minimal precision mode, while at the lowest precision there are no BG loops to unroll (i.e. the BG for-loop dimension size equals 1). 
  
  \subsection{Precision Scalability Modes}
   Existing SotA PSMA architectures can support different precision scalability modes. In the newly-introduced taxonomy, if the PSMA supports weight-only scalability, while the activations' precision is fixed at the highest precision, then the design is referred to as 1D precision scalable (e.g. 8b$\times$8b, 8b$\times$4b and 8b$\times$2b). 
   On the other hand, if both weight and activation precision are scalable, we denote this as 2D precision scalability. 2D scalability can be further classified to 2D symmetric (2D-S) scalability in case the precision of weights and activations are the same (e.g. 8b$\times$8b, 4b$\times$4b and 2b$\times$2b), or 2D asymmetric (2D-A) scalability in case the design supports different precisions for weights and activations (e.g. 8b$\times$8b, 8b$\times$4b, 8b$\times$2b, 4b$\times$4b and 2b$\times$2b).  
  
  
  \subsection{Fully Unrolled (FU) vs. Sub-Word Unrolled (SWU) Designs} \label{subsec:FU-SWU}
\begin{figure*}[!tb]
    \centering
    \includegraphics[width=\textwidth]{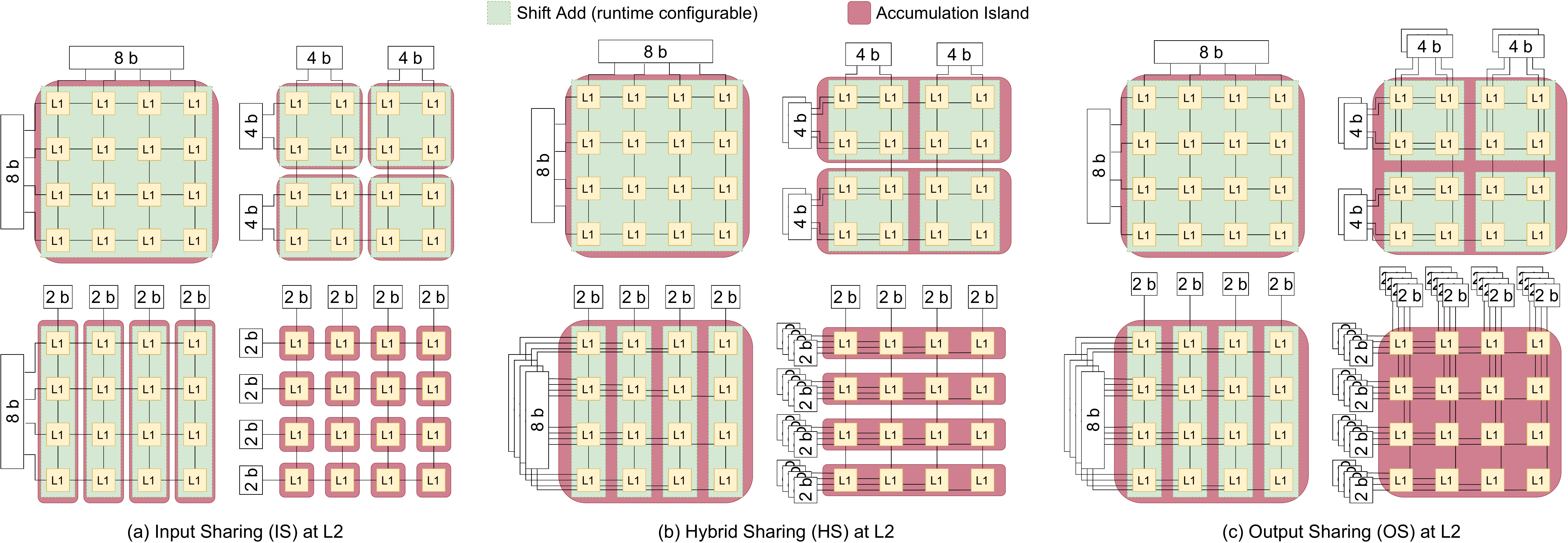}
    \caption{L2 in fully unrolled (FU) designs, when BG is unrolled in L2.}
    \label{fig:in_out_hybrid_share}
    \vspace{-1em}
  \end{figure*}

   As we move to lower precisions, a trade-off arises in terms of memory bandwidth vs. hardware utilization. To ensure full hardware utilization, the MAC array will require increased input/weight or output bandwidth when replacing the BG-loop unrolling with another loop unrolling. For instance, this can be seen in BitFusion \cite{sharma2018bitfusion} and BitBlade \cite{ryu2019bitblade} architectures, where the input/weight bandwidth is increased by a factor of 2x and 4x when scaling the precision down from 8-bit to 4-bit and 2-bit respectively. To illustrate this effect, \Cref{fig:in_out_hybrid_share} showcases how the introduced MAC array would behave at various precisions for different L2 unrolling schemes. For simplicity, only BG unrolling at L2 is shown, however, the same concepts and trade-offs apply for other BG configurations.
   
  At full precision, all unrolling schemes of \Cref{fig:in_out_hybrid_share} behave in the same way. That's because the BG loops are directly unrolled at that level, and there's no room to unroll any other loops. As we go down to lower precisions, the BG loops unrolling size reduces, thus enabling more unrolling of other for-loops. When unrolling an additional input sharing (IS) loop, the weights and activations can be broadcast across the vertical and horizontal dimensions. As a result, the input bandwidth is preserved. Yet, the outputs can not be added together, thus increasing the output bandwidth. On the other hand, when an output sharing (OS) loop is unrolled instead, all the partial results can be added together, preserving output bandwidth. Yet, all units should be fed with different inputs, hence requiring a higher input bandwidth. Hybrid sharing (HS) is the middle-ground between the two in terms of input/output bandwidth. As can be seen, the I/O bandwidth increases with lower precision operation for all designs.
  
\begin{figure}[!tb]
    \centering
    \includegraphics[width=0.48\textwidth]{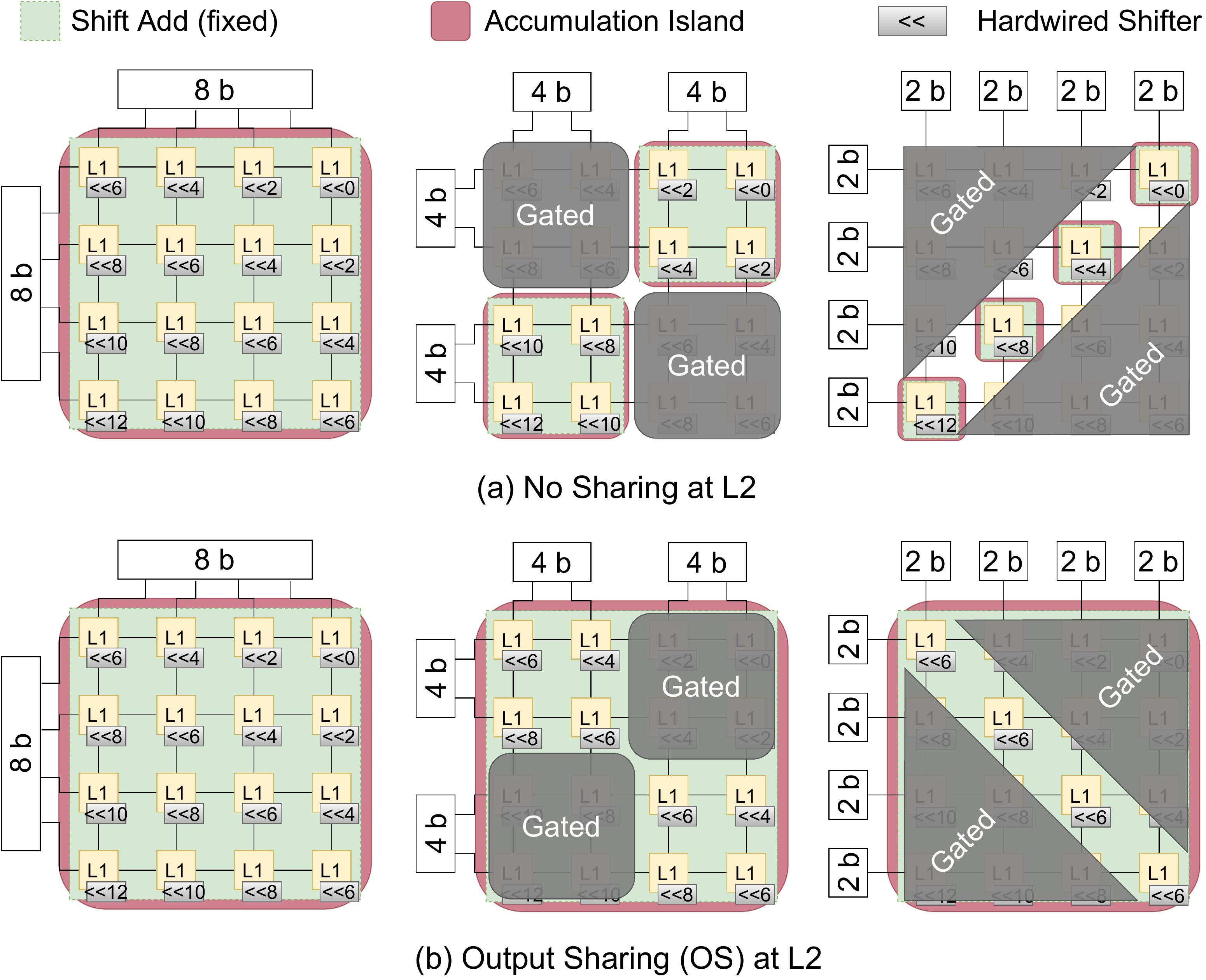}
    \caption{L2 in sub-word unrolled (SWU) designs, when BG is unrolled in L2.}
    \label{fig:SWU}
    \vspace{-1em}
  \end{figure}

   Some alternative PSMA topologies opt for a fixed I/O bandwidth across different precisions to simplify the memory interface design. In exchange, they opt for a lower hardware utilization at low precision, effectively gating part of the L1 units when scaling down in precision, as illustrated in \Cref{fig:SWU}. This family of MACs were first introduced in \cite{moons2017dvafs, mei2019subword}, and extended to a broader class of designs in \cite{camus2019review, camus2019survey}. We will here refer to them as sub-word unrolled (SWU) designs, in contrast to the fully unrolled (FU) topologies discussed earlier.
  
  \subsection{Complete Taxonomy and SotA Mapping}  
  Based on the concepts of previous subsections, we can now introduce the complete PSMA taxonomy. The new taxonomy is governed by 6 different parameters that fully define each architecture: 
  \begin{enumerate}
    \item \underline{L4/L3/L2:} Spatial unrolling at each level, selected from IS, HS or OS.
    \item \underline{BG:} Level at which the BG loops are unrolled, selected from spatially at L2/L3 or temporally in a bit-serial (BS) way. If BS, we also identify at which level the internal registers are located, denoted as BS-Lx.
    \item \underline{Config:} I/O bandwidth and hardware utilization tradeoff configuration, which can be FU or SWU.
    \item \underline{Mode:} Precision scalability, which can be 1D, 2D asymmetric, or 2D symmetric.
  \end{enumerate}
  
  This newly introduced taxonomy now allows to map all PSMA techniques presented in literature, as summarized in \Cref{tab:SotA_mapping}. Since BG unrolling is only relevant when the precision mode is above the minimal precision, a separate row per design is added for both high and low precisions.
  
\renewcommand{\arraystretch}{1.5} 
\setlength{\tabcolsep}{5pt} 

\begin{table}[!tb]
\centering
\caption{SotA Mapping}
\label{tab:SotA_mapping}
\begin{tabular}{@{}lccccccc@{}}
\toprule
SotA                                                   & Prec. & L4                                                                                       & L3                                                                                          & L2                                                                                          & BG                   & Config                 & Mode                   \\ \midrule
                                                       & High  & \cellBG{1}{double color fill={WS_Green}{IS_Blue}, shading angle=-45, opacity=1.0}{c}{IS} & \cellBG{1}{double color fill={OS_Red}{OS_Red}, shading angle=-45, opacity=1.0}{c}{OS}       & \cellBG{1}{double color fill={BG_Purple}{BG_Purple}, shading angle=-45, opacity=1.0}{c}{BG} &                      &                        &                        \\
\multirow{-2}{*}{DNPU \cite{shin2017dnpu}}             & Low   & \cellBG{1}{double color fill={WS_Green}{IS_Blue}, shading angle=-45, opacity=1.0}{c}{IS} & \cellBG{1}{double color fill={OS_Red}{OS_Red}, shading angle=-45, opacity=1.0}{c}{OS}       & \cellBG{1}{double color fill={WS_Green}{IS_Blue}, shading angle=-45, opacity=1.0}{c}{IS}    & \multirow{-2}{*}{L2} & \multirow{-2}{*}{FU} & \multirow{-2}{*}{1D}   \\ \midrule
                                                       & High  & \cellBG{1}{double color fill={WS_Green}{IS_Blue}, shading angle=-45, opacity=1.0}{c}{IS} & \cellBG{1}{double color fill={OS_Red}{OS_Red}, shading angle=-45, opacity=1.0}{c}{OS}       & \cellBG{1}{double color fill={BG_Purple}{BG_Purple}, shading angle=-45, opacity=1.0}{c}{BG} &                      &                        &                        \\
\multirow{-2}{*}{BitFusion \cite{sharma2018bitfusion}} & Low   & \cellBG{1}{double color fill={WS_Green}{IS_Blue}, shading angle=-45, opacity=1.0}{c}{IS} & \cellBG{1}{double color fill={OS_Red}{OS_Red}, shading angle=-45, opacity=1.0}{c}{OS}       & \cellBG{1}{double color fill={OS_Red}{OS_Red}, shading angle=-45, opacity=1.0}{c}{OS}       & \multirow{-2}{*}{L2} & \multirow{-2}{*}{FU} & \multirow{-2}{*}{2D-A} \\ \midrule
                                                       & High  & \cellBG{1}{double color fill={WS_Green}{IS_Blue}, shading angle=-45, opacity=1.0}{c}{IS} & \cellBG{1}{double color fill={BG_Purple}{BG_Purple}, shading angle=-45, opacity=1.0}{c}{BG} & \cellBG{1}{double color fill={OS_Red}{OS_Red}, shading angle=-45, opacity=1.0}{c}{OS}       &                      &                        &                        \\
\multirow{-2}{*}{BitBlade \cite{ryu2019bitblade}}      & Low   & \cellBG{1}{double color fill={WS_Green}{IS_Blue}, shading angle=-45, opacity=1.0}{c}{IS} & \cellBG{1}{double color fill={OS_Red}{OS_Red}, shading angle=-45, opacity=1.0}{c}{OS}       & \cellBG{1}{double color fill={OS_Red}{OS_Red}, shading angle=-45, opacity=1.0}{c}{OS}       & \multirow{-2}{*}{L3} & \multirow{-2}{*}{FU} & \multirow{-2}{*}{2D-A} \\ \midrule
                                                       & High  & \cellBG{1}{double color fill={WS_Green}{IS_Blue}, shading angle=-45, opacity=1.0}{c}{IS} & \cellBG{1}{double color fill={BG_Purple}{BG_Purple}, shading angle=-45, opacity=1.0}{c}{BG} & \cellBG{1}{double color fill={OS_Red}{OS_Red}, shading angle=-45, opacity=1.0}{c}{OS}       &                      &                        &                        \\
\multirow{-2}{*}{Ghodrati \cite{ghodrati2020bit}}      & Low   & \cellBG{1}{double color fill={WS_Green}{IS_Blue}, shading angle=-45, opacity=1.0}{c}{IS} & \cellBG{1}{double color fill={OS_Red}{OS_Red}, shading angle=-45, opacity=1.0}{c}{OS}       & \cellBG{1}{double color fill={OS_Red}{OS_Red}, shading angle=-45, opacity=1.0}{c}{OS}       & \multirow{-2}{*}{L3} & \multirow{-2}{*}{FU} & \multirow{-2}{*}{2D-A} \\ \midrule
                                                       & High  & \cellBG{1}{double color fill={WS_Green}{IS_Blue}, shading angle=-45, opacity=1.0}{c}{IS} & \cellBG{1}{double color fill={WS_Green}{IS_Blue}, shading angle=-45, opacity=1.0}{c}{IS}    & \cellBG{1}{double color fill={OS_Red}{OS_Red}, shading angle=-45, opacity=1.0}{c}{OS}       &                      &                        &                        \\
\multirow{-2}{*}{Stripes \cite{judd2016stripes}}       & Low   & \cellBG{1}{double color fill={WS_Green}{IS_Blue}, shading angle=-45, opacity=1.0}{c}{IS} & \cellBG{1}{double color fill={WS_Green}{IS_Blue}, shading angle=-45, opacity=1.0}{c}{IS}    & \cellBG{1}{double color fill={OS_Red}{OS_Red}, shading angle=-45, opacity=1.0}{c}{OS}       & \multirow{-2}{*}{BS-L2} & \multirow{-2}{*}{FU} & \multirow{-2}{*}{1D}   \\ \midrule
                                                       & High  & \cellBG{1}{double color fill={WS_Green}{IS_Blue}, shading angle=-45, opacity=1.0}{c}{IS} & \cellBG{1}{double color fill={OS_Red}{OS_Red}, shading angle=-45, opacity=1.0}{c}{OS}       & \cellBG{1}{double color fill={WS_Green}{IS_Blue}, shading angle=-45, opacity=1.0}{c}{IS}    &                      &                        &                        \\
\multirow{-2}{*}{UNPU \cite{lee2018unpu}}              & Low   & \cellBG{1}{double color fill={WS_Green}{IS_Blue}, shading angle=-45, opacity=1.0}{c}{IS} & \cellBG{1}{double color fill={OS_Red}{OS_Red}, shading angle=-45, opacity=1.0}{c}{OS}       & \cellBG{1}{double color fill={WS_Green}{IS_Blue}, shading angle=-45, opacity=1.0}{c}{IS}    & \multirow{-2}{*}{BS-L1} & \multirow{-2}{*}{FU} & \multirow{-2}{*}{1D}   \\ \midrule
                                                       & High  & \cellBG{1}{double color fill={WS_Green}{IS_Blue}, shading angle=-45, opacity=1.0}{c}{IS} & \cellBG{1}{double color fill={WS_Green}{IS_Blue}, shading angle=-45, opacity=1.0}{c}{IS}    & \cellBG{1}{double color fill={OS_Red}{OS_Red}, shading angle=-45, opacity=1.0}{c}{OS}       &                      &                        &                        \\
\multirow{-2}{*}{Loom \cite{sharify2018loom}}          & Low   & \cellBG{1}{double color fill={WS_Green}{IS_Blue}, shading angle=-45, opacity=1.0}{c}{IS} & \cellBG{1}{double color fill={WS_Green}{IS_Blue}, shading angle=-45, opacity=1.0}{c}{IS}    & \cellBG{1}{double color fill={OS_Red}{OS_Red}, shading angle=-45, opacity=1.0}{c}{OS}       & \multirow{-2}{*}{BS-L2} & \multirow{-2}{*}{FU} & \multirow{-2}{*}{2D-A} \\ \midrule
                                                       & High  & \cellBG{1}{double color fill={WS_Green}{IS_Blue}, shading angle=-45, opacity=1.0}{c}{IS} & \cellBG{1}{double color fill={WS_Green}{IS_Blue}, shading angle=-45, opacity=1.0}{c}{IS}    & \cellBG{1}{double color fill={BG_Purple}{BG_Purple}, shading angle=-45, opacity=1.0}{c}{BG} &                      &                        &                        \\
\multirow{-2}{*}{Envision \cite{moons2017envision}}    & Low   & \cellBG{1}{double color fill={WS_Green}{IS_Blue}, shading angle=-45, opacity=1.0}{c}{IS} & \cellBG{1}{double color fill={WS_Green}{IS_Blue}, shading angle=-45, opacity=1.0}{c}{IS}    & \cellBG{1}{double color fill={Grey}{Grey}, shading angle=-45, opacity=1.0}{c}{No}         & \multirow{-2}{*}{L2} & \multirow{-2}{*}{SWU}  & \multirow{-2}{*}{2D-S} \\ \midrule
                                                       & High  & \cellBG{1}{double color fill={WS_Green}{IS_Blue}, shading angle=-45, opacity=1.0}{c}{IS} & \cellBG{1}{double color fill={OS_Red}{OS_Red}, shading angle=-45, opacity=1.0}{c}{OS}       & \cellBG{1}{double color fill={BG_Purple}{BG_Purple}, shading angle=-45, opacity=1.0}{c}{BG} &                      &                        &                        \\
\multirow{-2}{*}{ST \cite{mei2019subword}}         & Low   & \cellBG{1}{double color fill={WS_Green}{IS_Blue}, shading angle=-45, opacity=1.0}{c}{IS} & \cellBG{1}{double color fill={OS_Red}{OS_Red}, shading angle=-45, opacity=1.0}{c}{OS}       & \cellBG{1}{double color fill={OS_Red}{OS_Red}, shading angle=-45, opacity=1.0}{c}{OS}       & \multirow{-2}{*}{L2} & \multirow{-2}{*}{SWU}  & \multirow{-2}{*}{2D-S} \\ \midrule[\heavyrulewidth]
\multicolumn{8}{l}{\footnotesize \textbf{IS:} Input\&Weight Sharing, \textbf{OS:} Output Sharing, \textbf{HS:} Hybrid Sharing,}  \\ 
\multicolumn{8}{l}{\footnotesize \textbf{BS-Lx:} Bit-Serial with internal shift-add registers located at level Lx, }   
  \\ 
\multicolumn{8}{l}{\footnotesize \textbf{SWU:} Sub-Word Unrolled, \textbf{FU:} Fully Unrolled,}   
\\ 
\multicolumn{8}{l}{\footnotesize \textbf{1D:} 1D scalability, \textbf{2D-A/S:} 2D Asymmetric/Symmetric scalability.}                                      \\
\bottomrule
\end{tabular}
\vspace{-1em}
\end{table}

  For DNPU \cite{shin2017dnpu} and BitFusion \cite{sharma2018bitfusion}, each L2 unit produces a full product every clock cycle, thus their BG are unrolled spatially at L2. The differences between them is that BitFusion is 2D asymmetric scalable, while DNPU is only 1D scalable. Additionally, at low precisions, BitFusion's L2 is output shared, while DNPU's L2 is input/weight shared. BitBlade \cite{ryu2019bitblade} and Ghodrati \cite{ghodrati2020bit} have a very similar PSMA architecture, and thus are mapped the same way in the new taxonomy. In their works, the shifters are shared between L2 units at L3, thus their BG is unrolled spatially at L3. 
  
  Stripes \cite{judd2016stripes}, UNPU \cite{lee2018unpu}, and Loom \cite{sharify2018loom} are bit-serial designs, meaning the BG loops are unrolled temporally rather than spatially. Stripes and UNPU are 1D scalable, while Loom is 2D scalable. Additionally, UNPU is IS at L2 with its internal shift-add registers located at L1, while Stripes and Loom are OS at L2 with its internal registers located at L2. 
  
  The final batch include Envision \cite{moons2017envision} and ST \cite{mei2019subword}, which are both SWU designs. Fundamentally, the difference between them is their behaviour at L2 in low precisions, where Envision has no sharing, where ST is OS. Moreover, Envision opts for an IS scheme at L3, while ST goes for an OS scheme.
  
  With the introduction of the new taxonomy, the road is paved towards benchmarking the full design space under the same operating circumstances and technology, leading to better insights on the pros and cons of each PSMA architecture. To achieve this goal, a uniform and highly parameterizable MAC array template is built, which allows to efficiently map different design points covered by the taxonomy. To the best of our knowledge, this work is the first attempt to qualitatively compare between different PSMA architectures.
  
  \section{Uniform and Parameterized PSMA Template} \label{sec:uniform_mac}
  
  The SotA implementations presented in \Cref{tab:SotA_mapping} are not the only possible PSMA instantiations. The taxonomy allows to identify a broader range of possible architectures that were not previously covered in literature yet. To be able to quickly implement and benchmark all different configurations in the introduced design space, a uniform and highly parameterized PSMA template is developed. Based on its user-defined parameters, this flexible PSMA can be design-time configured into any array architecture in a subset of the design space spanned by the introduced taxonomy, i.e. it supports different L2, L3, L4, BG, and FU/SWU settings.
  
  \subsection{Bit-Group Configurations}
\begin{figure*}[!tb]
    \centering
    \includegraphics[width=0.95\textwidth]{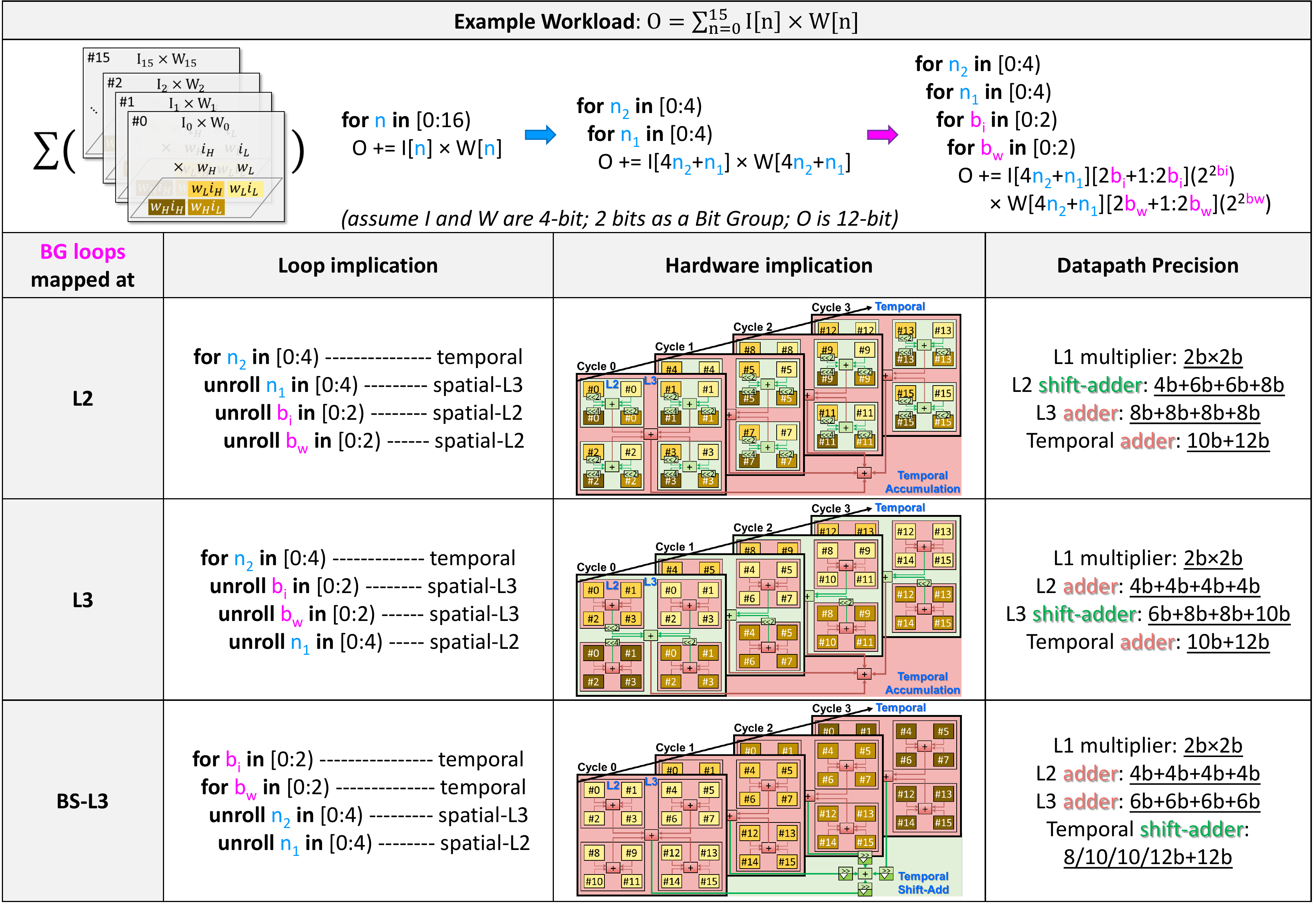}
    \caption{BG is unrolled spatially at L2, L3, or temporally at L3 (BS-L3). Assume OS at L2 and L3; Assume each level $L_n$ contains 2$\times$2 $L_{n-1}$ for simplicity.}
    \vspace{-1em}
    \label{fig:BG_unrolling}
  \end{figure*}

  The uniform parameterized PSMA template supports different BG unrollings. The HW implication of this is illustrated in \Cref{fig:BG_unrolling} in a simplified way. At full precision, if BG is unrolled spatially at L2, it means that each L2 unit produces a full product. It also means that the shift \& add tree lives inside each one of the L2 units, and thus the shifters are not amortized across L1 results as in the case of L3 and BS BG unrollings. 
  
  On the other hand, if BG loops are spatially unrolled at L3, it means that the shift \& add tree lives in L3 instead of L2, and are shared between aggregated L1 results. To guarantee correct functionality, each L2 unit only multiplies bit-groups that share the same significance, as each L2 product will be shifted by the same amount. To illustrate, in \Cref{fig:BG_unrolling}, if BG loops are unrolled at L3, you can find that all the BG-level operations with the same color are shifted by the same amount, and thus grouped in the same L2 unit.
  
  For bit-serial designs, the BGs are unrolled temporally rather than spatially. This means that the adders and shifters are de-coupled, and the shifting operation is performed over multiple clock cycles, depending on the precision. Additional circuitry is required to ensure correct functionality of BS designs, such as timer logic for scheduling and internal registers. It's worth noting that by default, only BS designs include internal registers between array levels.
  
  \begin{figure}[!tb]
    \centering
    \includegraphics[width=3.2in]{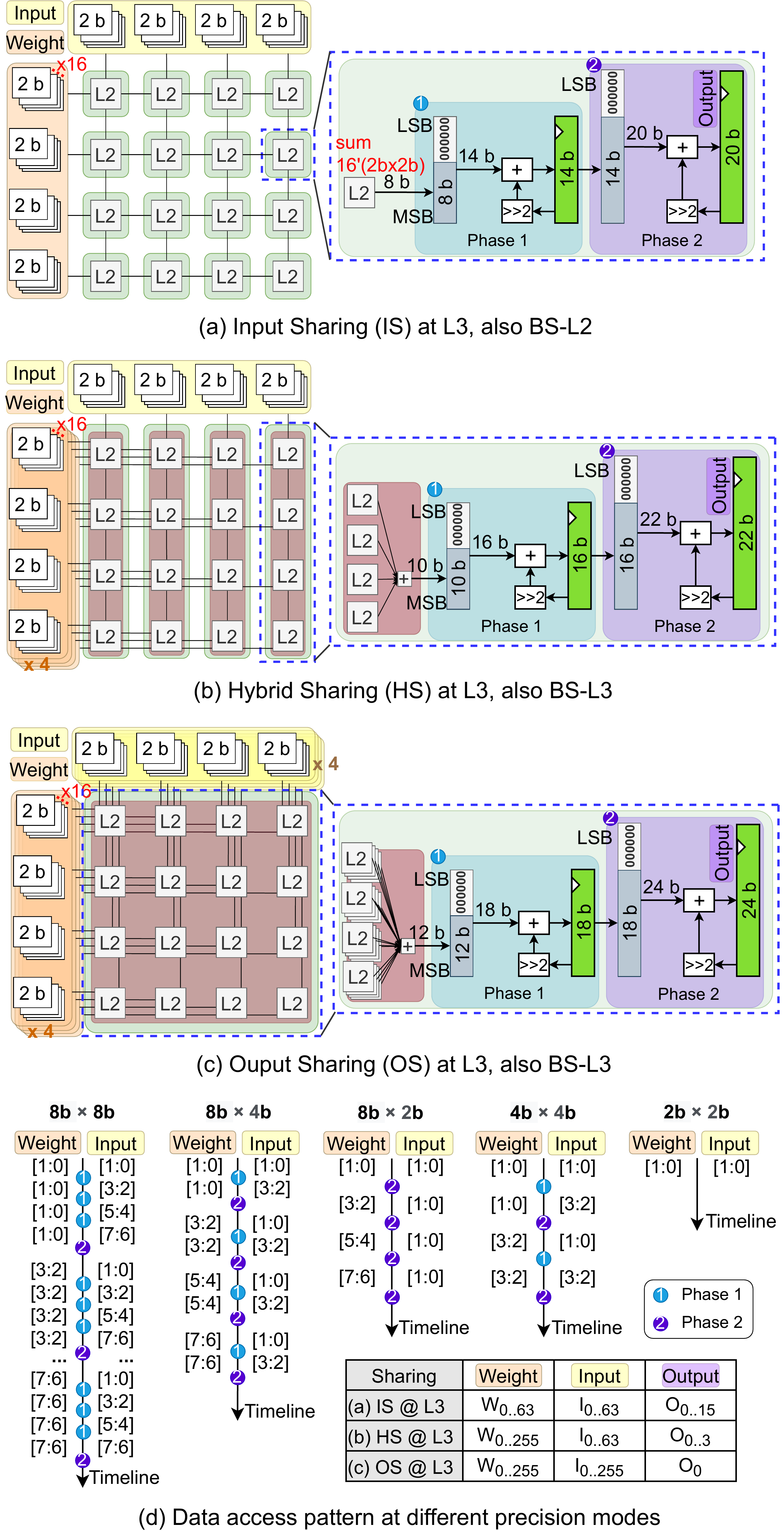}
    \caption{Input, Hybrid, and Output Sharing at L3 for different precisions when BG is unrolled temporally. Assume OS at L2.}
    \label{fig:temporal_BG}
  \end{figure}
  
  To gain a better understanding of the internal architecture of bit-serial array designs and how the scheduling is handled, refer to \Cref{fig:temporal_BG}. Assuming L2 units are OS, then each L2 would produce one 8-bits result (16 2b$\times$2b results sum together). The scheduling is done in two phases. First, the weights are stationary while the inputs are shifted right by 2-bits each clock cycle (Phase 1). During that time, the intermediate results are stored in the first internal register. After all the input bits are depleted, the weights are shifted right by 2-bits, the data stored in the first internal register is transferred to the second internal register (Phase 2), and then Phase 1 is repeated again. This cycle repeats itself until all the weight bits are depleted, then finally the accumulator at the end of the array (not shown in \Cref{fig:temporal_BG}) gets activated, and accumulates the full product. 
  
  As you may have noticed in \Cref{fig:BG_unrolling}, bit-serial designs reduce the complexity of the L2 and L3 adder trees, as they don't require configurable shifters anymore. On the other hand, bit-serial designs have their own limitations. One is that they have to find enough other for-loops (other than BG loops) in the algorithm that can be spatially unrolled on the array to ensure high hardware utilization. Secondly, bit-serial designs require additional scheduling control logic and internal circuitry for each partial output. It is hence beneficial to reduce the number of outputs generated when design in this pattern, especially at lower levels of the array. With that in mind, lots of inefficient designs can already be pruned out from the full design space. 
  
  \subsection{Design Space Constraints}\label{subsec:constraints}

\renewcommand{\arraystretch}{1.2} 
\setlength{\tabcolsep}{3.5pt} 

\begin{table}[!tb]
\centering
\caption{Constrained Design Space}
\label{tab:all_mappings}
\begin{tabular}{@{}lcccccccc@{}}
\toprule
Layer & \multicolumn{8}{c}{Supported Configurations}                                     \\ \midrule
L4    & \cellBG{3}{double color fill={WS_Green}{IS_Blue}, shading angle=-45, opacity=1.0}{c}{IS} & \cellBG{2}{double color fill={OS_Red}{IS_Blue}, shading angle=-45, opacity=1.0}{c}{HS}         & \cellBG{3}{double color fill={OS_Red}{OS_Red}, shading angle=-45, opacity=1.0}{c}{OS} \\ \cmidrule(lr){2-9}
L3    & \cellBG{3}{double color fill={WS_Green}{IS_Blue}, shading angle=-45, opacity=1.0}{c}{IS} & \cellBG{2}{double color fill={OS_Red}{IS_Blue}, shading angle=-45, opacity=1.0}{c}{HS}         & \cellBG{3}{double color fill={OS_Red}{OS_Red}, shading angle=-45, opacity=1.0}{c}{OS} \\ \cmidrule(lr){2-9}
Config   & \multicolumn{6}{c}{FU}                                 & \multicolumn{2}{c}{SWU} \\ \cmidrule(lr){2-7} \cmidrule(lr){8-9}
BG    & \multicolumn{3}{c}{L2} & \multicolumn{2}{c}{L3} & BS-L2    & \multicolumn{2}{c}{L2} \\ \cmidrule(lr){2-4} \cmidrule(lr){5-6} \cmidrule(lr){7-7} \cmidrule(lr){8-9}
L2    & \cellBG{1}{double color fill={WS_Green}{IS_Blue}, shading angle=-45, opacity=1.0}{c}{IS}     & \cellBG{1}{double color fill={OS_Red}{IS_Blue}, shading angle=-45, opacity=1.0}{c}{HS}    & \cellBG{1}{double color fill={OS_Red}{OS_Red}, shading angle=-45, opacity=1.0}{c}{OS}    & \cellBG{1}{double color fill={OS_Red}{IS_Blue}, shading angle=-45, opacity=1.0}{c}{HS}         & \cellBG{1}{double color fill={OS_Red}{OS_Red}, shading angle=-45, opacity=1.0}{c}{OS}        & \cellBG{1}{double color fill={OS_Red}{OS_Red}, shading angle=-45, opacity=1.0}{c}{OS}    & \cellBG{1}{double color fill={WS_Green}{IS_Blue}, shading angle=-45, opacity=1.0}{c}{IS}         & \cellBG{1}{double color fill={OS_Red}{OS_Red}, shading angle=-45, opacity=1.0}{c}{OS}        \\ \midrule[\heavyrulewidth]
\multicolumn{9}{l}{\footnotesize \textbf{Number of supported designs:}}  \\ 
\multicolumn{9}{l}{\footnotesize $3\textrm{ (L4)} \times 3\textrm{ (L3)} \times 8\textrm{ (Config / BG / L2)} = 72$ designs}     \\ \bottomrule
\end{tabular}
\vspace{-1.5em}
\end{table}
  
  Based on the taxonomy introduced in \Cref{sec:taxonomy}, the combination of the design parameters would lead to a huge variety of array configurations. As such, some constraints had to be put in place to reduce the design space to make it more manageable, while still maintaining the interesting exploration regions. With the observations of previous subsections, the list below summarizes the constraints that were imposed on the different parameters, and their justification. The constrained design space is also summarized in \Cref{tab:all_mappings}.
  \begin{enumerate}
      \item \underline{If BG is unrolled at L3, L2 cannot be IS.} The benefit of BG in L3 is that all the L1 units within L2 can share shifting logic. If at L2 the outputs are not shared, we get the same number of shifters as if BG is unrolled in L2. 
      \item \underline{If BG is temporally unrolled, L2 is OS with BS-L2.} BS designs require internal registers to support the temporal shifting of each output. In this work, we fixed the location of these BS internal registers to L2, to reduce the template complexity and balance the critical path. To minimize their overhead, it is best if only a single internal register per L2 unit is needed, resulting in a total of 256 internal registers. BS-L3/L4's benchmarking can be interesting as well, but would lead to many more input registers and requires large OS in the workload. This study is left to future work.
      \item \underline{If SWU design, L2 can either be OS or have no sharing.} Due to SWU designs' nature, L2 have to be symmetrical. Thus, the outputs are either shared or not shared. 
  \end{enumerate}
  
  It's worth noting that, given the design constraints, UNPU \cite{lee2018unpu} is the only accelerator in the SotA mapping that is not supported by our uniform PSMA template, given that it's a BS design with L2 being IS (BS-L1).

  \subsection{Register Layout}
  
  \begin{figure}[!tb]
    \centering
    \includegraphics[width=\columnwidth]{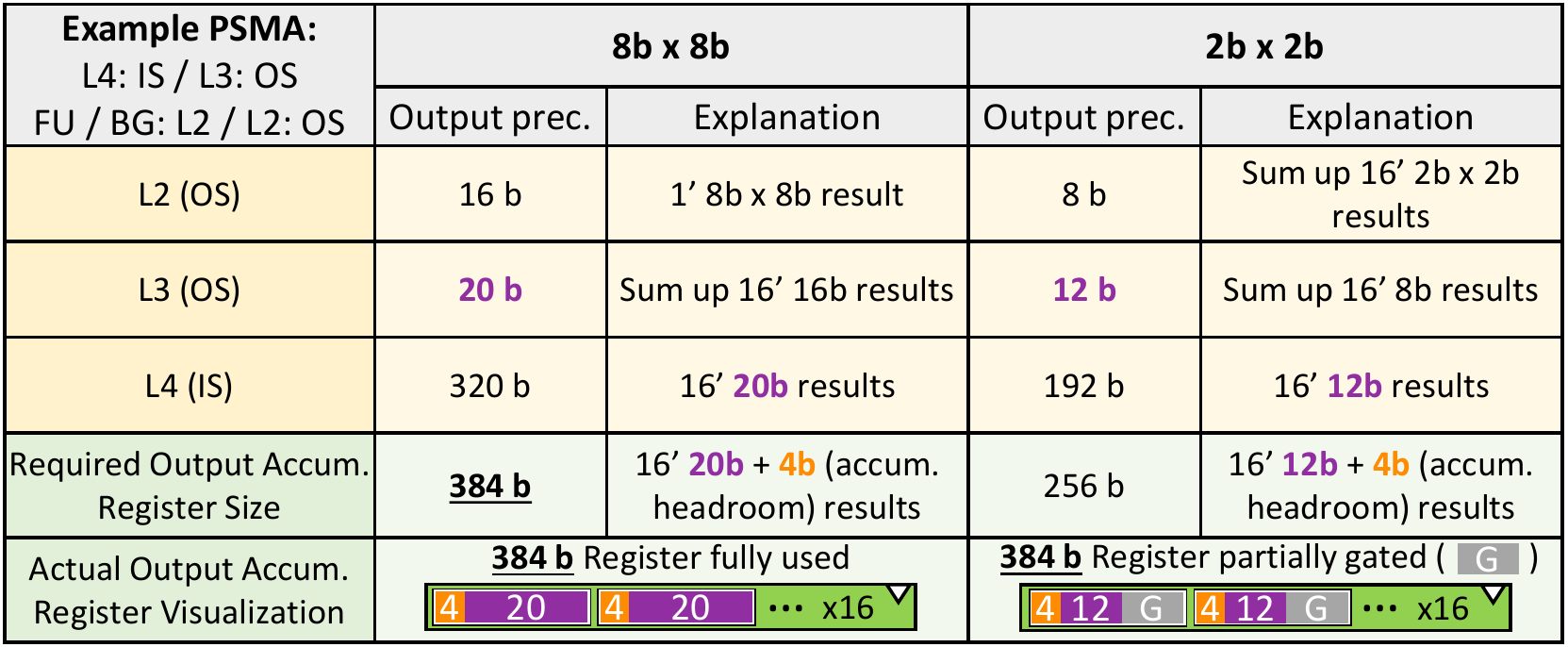}
    \caption{An example of output accumulator register size calculation and its run-time configurations over full/low precision modes.}
    \label{fig:output_reg_vis}
  \end{figure}
  
  To ensure a proper timing of all designs assessed using the uniform PSMA template, we include input and output registers at the periphery of the array template. Only the BS designs do include additional internal registers within the array to assist periodic shift-add process. Depending on the array L4/L3/L2 configurations (namely, the IS/OS configuration as well as the BG unrolling, as discussed in \Cref{sec:taxonomy}), each design requires a different maximum input/output bandwidth per clock cycle. As a result, the required amount of registers at the array's inputs/output will vary. IS/WS designs typically require less input registers, at the expense of more output registers. The opposite is true for OS schemes.
  
  Practically, each OS / IS sharing dimension along each hierarchical levels L2/L3/L4 multiplies the required number of input / output register words by 4 (since each level $L_{n}$ is a 2-dimensional 4$\times$4 $L_{n-1}$ array). It is hereby important to note that the number of bits per input word is fixed (8-bit at full precision), while the number of bits per output word depends both on the input word precision and the maximal expected temporal accumulation time. Here, the largest required bit width across all precision modes is computed (take the worst case scenario), with an extra 4-bits of headroom for temporal accumulation per each expected output. An example of how we compute the final output register size is given in \Cref{fig:output_reg_vis}, in which, for an example PSMA architecture, the required output precision after each level (from L2 to L4, further to accumulator) are given and explained.
  
  Finally, two-stage accumulation registers are needed in BS designs. Since the design space has been constrained in \Cref{subsec:constraints} to only include BS-L2, the BS registers layout is always the same. As shown in \Cref{fig:temporal_BG}(a), L2 in BS designs always produce 8-bit results, as it accumulates 16 2$\times$2-bit partial products. At full precision, the internal registers clear their stored value every 4 iterations and each register's stored value needs to be shifted right by 2 for 3 times, therefore a 6-bit headroom is needed for each BS accumulation register.

  \begin{figure*}[t]
    \centering
    \subfloat[200 MHz]{
      \includegraphics[height=7.1cm]{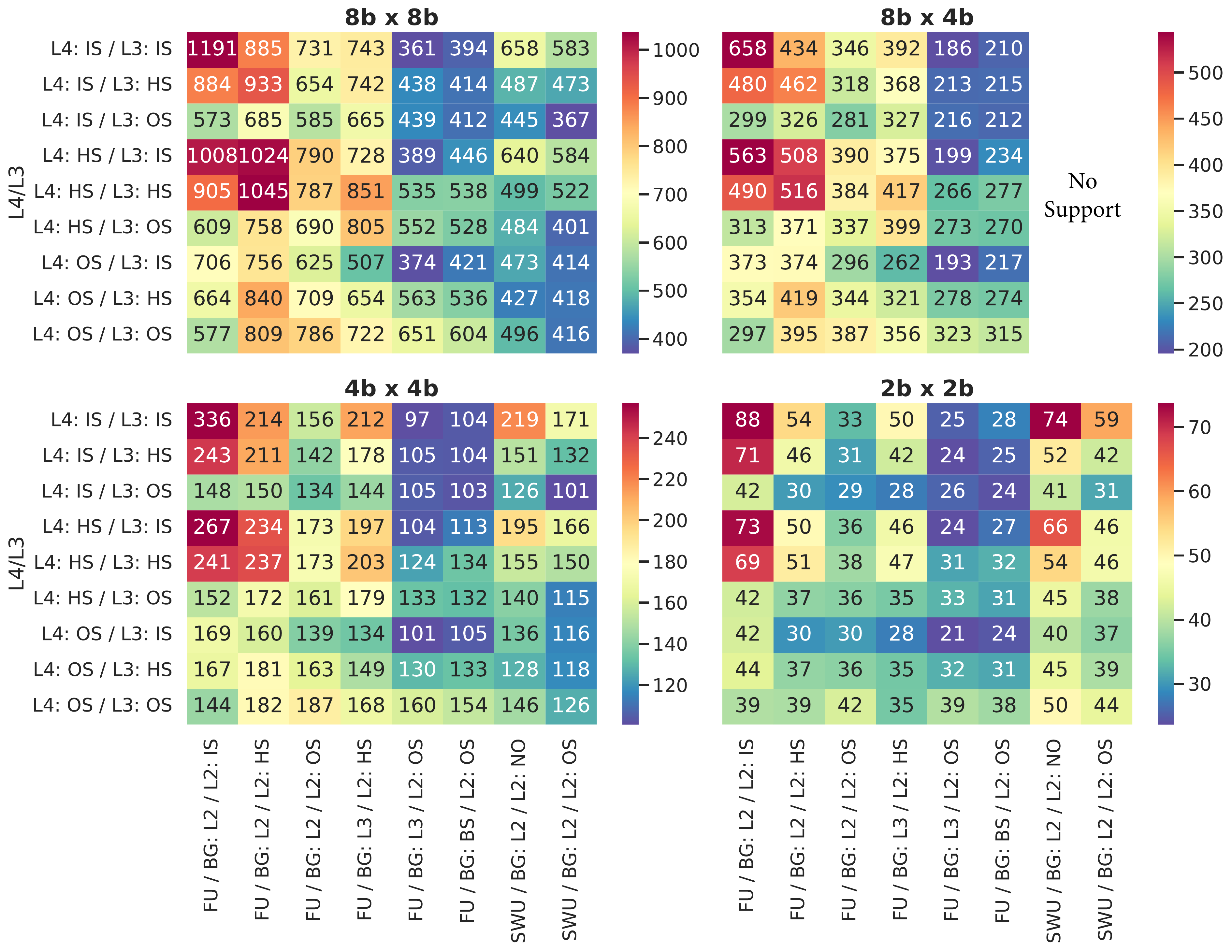}
      \label{subfig:Heatmap-0.2}
    }
    \subfloat[1 GHz]{
      \includegraphics[height=7.1cm]{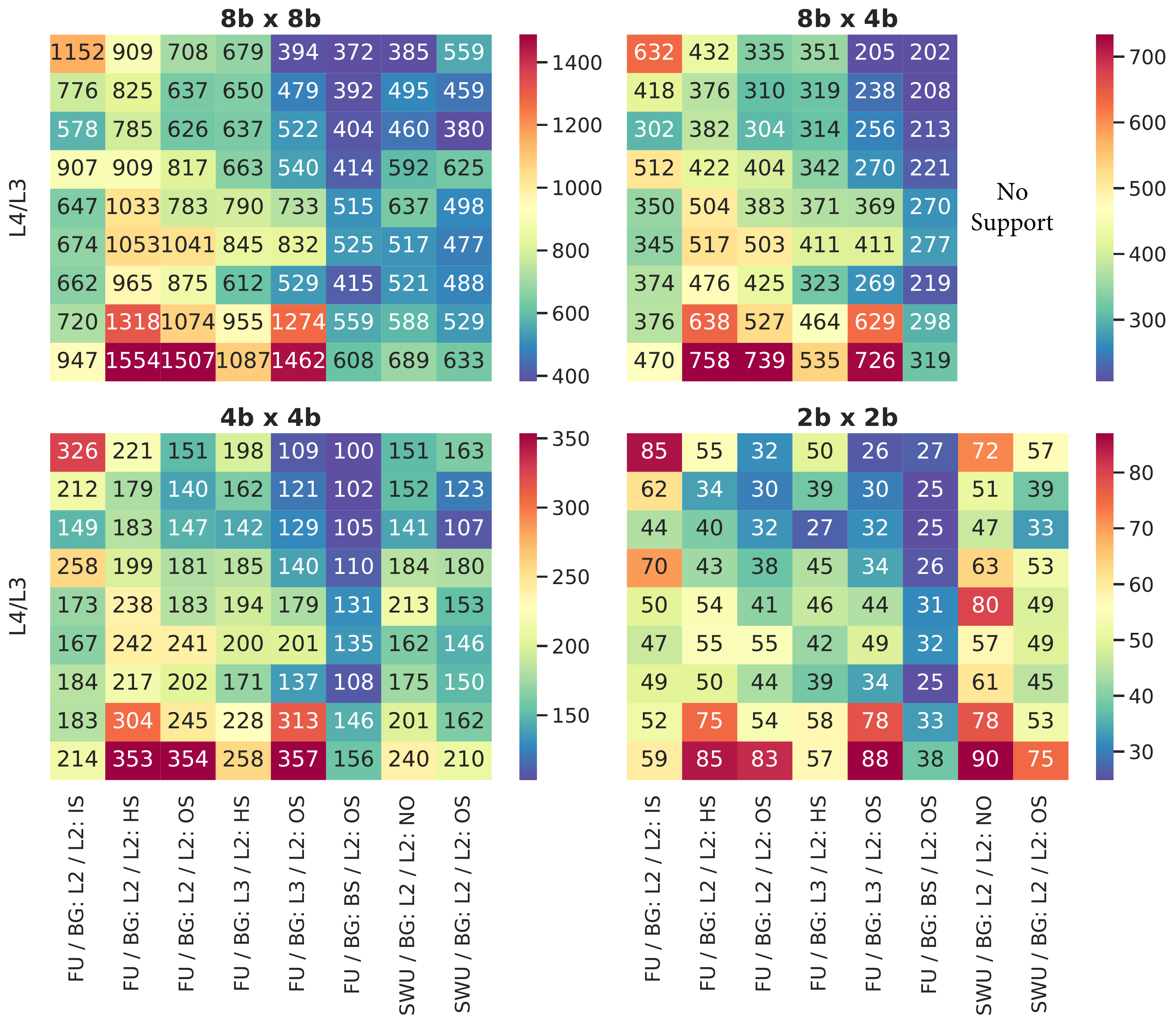}
      \label{subfig:Heatmap-1}
    }
    \caption{Energy per Operation ($fJ$) heatmaps at (a) 200 MHz, and (b) 1 GHz.}
    \label{fig:Heatmap-all}
  \end{figure*}

\section{Experimental Setup} \label{sec:comparative_study}
  In this section, each design in the design space is evaluated in terms of energy per operation and area, both in a low frequency (200MHz) and high frequency (1GHz) context. Additionally, breakdowns are performed to gain more insights on the trade-off between hardware components of systems with different design-time and run-time configurations. 
  
  \subsection{Methodology}
  To ensure a fair comparison, all 72 benchmarked designs are synthesized from the same parameterizable HDL template with the same coding style and set of optimizations. For the precision configuration, most designs can handle both symmetrical and weight-only scaling, with the exception of SWU designs that only support symmetrical precision scalability. Supported precision scaling values are 8-bits, 4-bits, and 2-bits for both input activations and weights. 
  
  All designs are synthesized from SystemVerilog HDL using Cadence Genus v19.11 with high-effort compilation, once at 200MHz, and once at 1GHz. The used technology node is TSMC 28-nm, with a nominal supply voltage of 1V. Power estimations are conducted through post-synthesis simulations, using Questa Sim v10.6c for simulating switching activity and Cadence Genus for power extraction. The simulations were conducted for 4,096 clock cycles with an ideal workload, one that ensures full hardware utilization for all designs at all precision modes. BS designs benchmarked in this section are all BS-L2. For simplicity, we refer them directly as BS.
  
  \subsection{Workload}
  To further elaborate on the nature of the ideal workload, it can be characterized in terms of CNN for-loops. At lowest precision (2b$\times$2b), the PSMA maximally consist of 4,096 functional PEs. So, the workload should at least have 4,096 OS loops and 4,096 IS/WS loops to guarantee full hardware utilization for all designs at all precisions. \Cref{tab:workloads} shows a breakdown how the number of for-loops was chosen for each loop category, both for all-level IS and all-level OS corner cases. Basically, all-level IS designs will spatially unroll IS/WS loops of the workload, while OS loops will be unrolled temporally, thus providing more output stationarity. The opposite is true for all-level OS designs. Ultimately, all designs will run the same workload with randomly generated inputs and weights, but with different dataflows and mappings.
  
  \subsection{Throughput}
  All simulations are executing an optimized workload for which all designs have a spatial utilization of 100\% at nominal precision. Hence, both FU and SWU designs have the same throughput (number of operations per clock cycle) at full precision. As we scale down to lower precisions, the FU designs still utilize the full hardware, while SWU gate part of the PSMA in order to maintain fixed input bandwidth across all precisions. As a result, SWU has reduced throughput compared to FU, specifically, half and quarter the throughput at 4-bit and 2-bit precisions respectively.
  
\renewcommand{\arraystretch}{1.5} 
\setlength{\tabcolsep}{5pt} 

\begin{table}[!tb]
\vspace{-1em}
\centering
\caption{Ideal workload*'s minimal size}
\label{tab:workloads}
\begin{tabular}{@{}lccc@{}}
\toprule
Corner cases at 2b$\times$2b    & IS loops & WS loops & OS loops \\ \midrule
\cellBG{1}{double color fill={WS_Green}{IS_Blue}, shading angle=-45, opacity=1.0}{c}{L4, L3, L2: IS}                  & 64       & 64       & 1        \\
\cellBG{1}{double color fill={OS_Red}{OS_Red}, shading angle=-45, opacity=1.0}{c}{L4, L3, L2: OS}                  & 1        & 1        & 4096     \\ \midrule[\heavyrulewidth]
Overall required loop size                           & 64       & 64       & 4096     \\ \bottomrule
\multicolumn{4}{l}{\footnotesize \textbf{*} Workload that guarantees full hardware utilization} \\ 
\multicolumn{4}{l}{\footnotesize \hspace{5pt} for all designs across all precisions.}
\end{tabular}
\vspace{-1em}
\end{table}
  
    \begin{figure*}[!tb]
    \centering
    \subfloat[200 MHz]{
      \includegraphics[height=7.5cm]{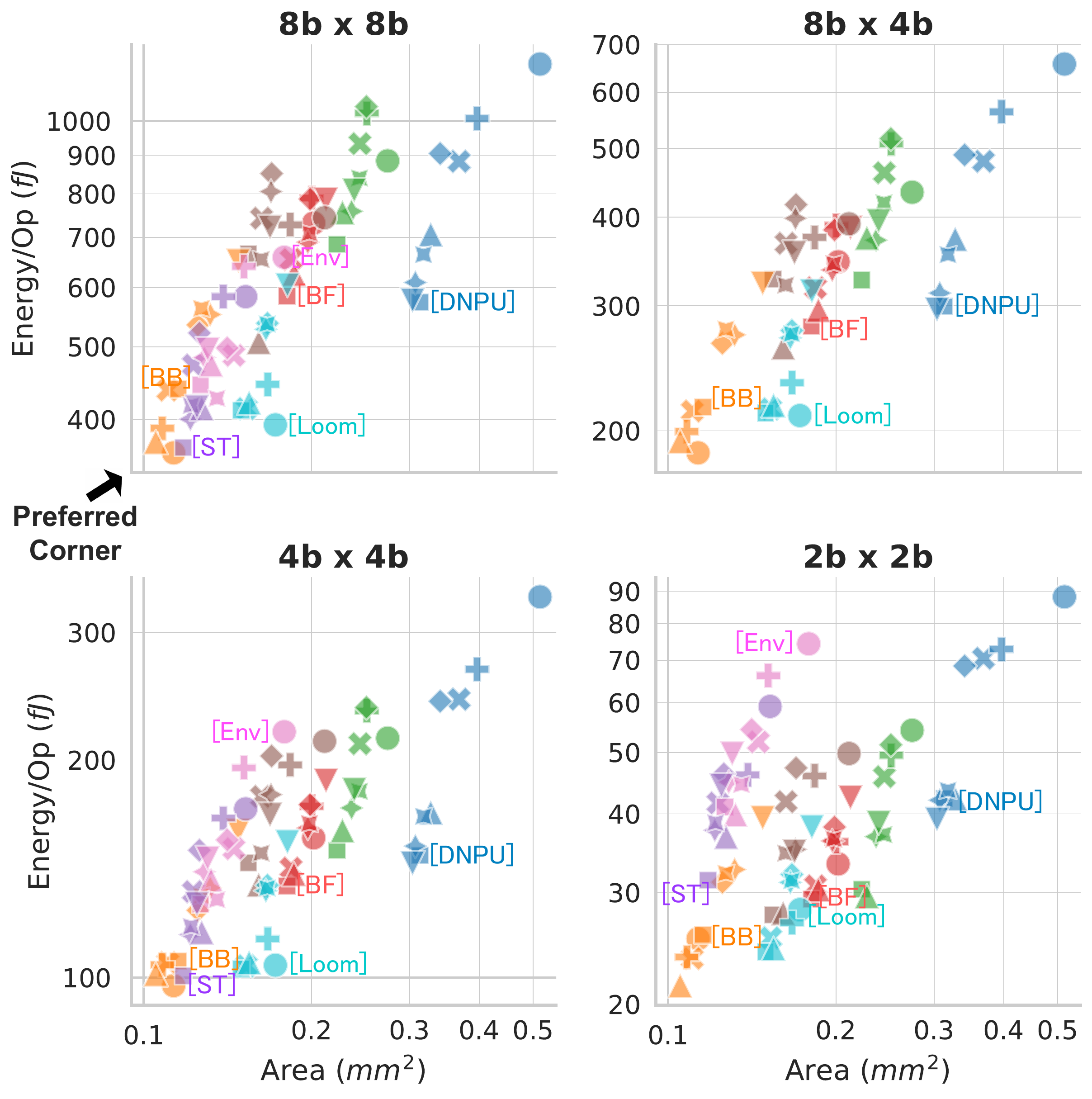}
      \label{subfig:Scatter-0.2}
    }
    \subfloat[1 GHz]{
      \includegraphics[height=7.5cm]{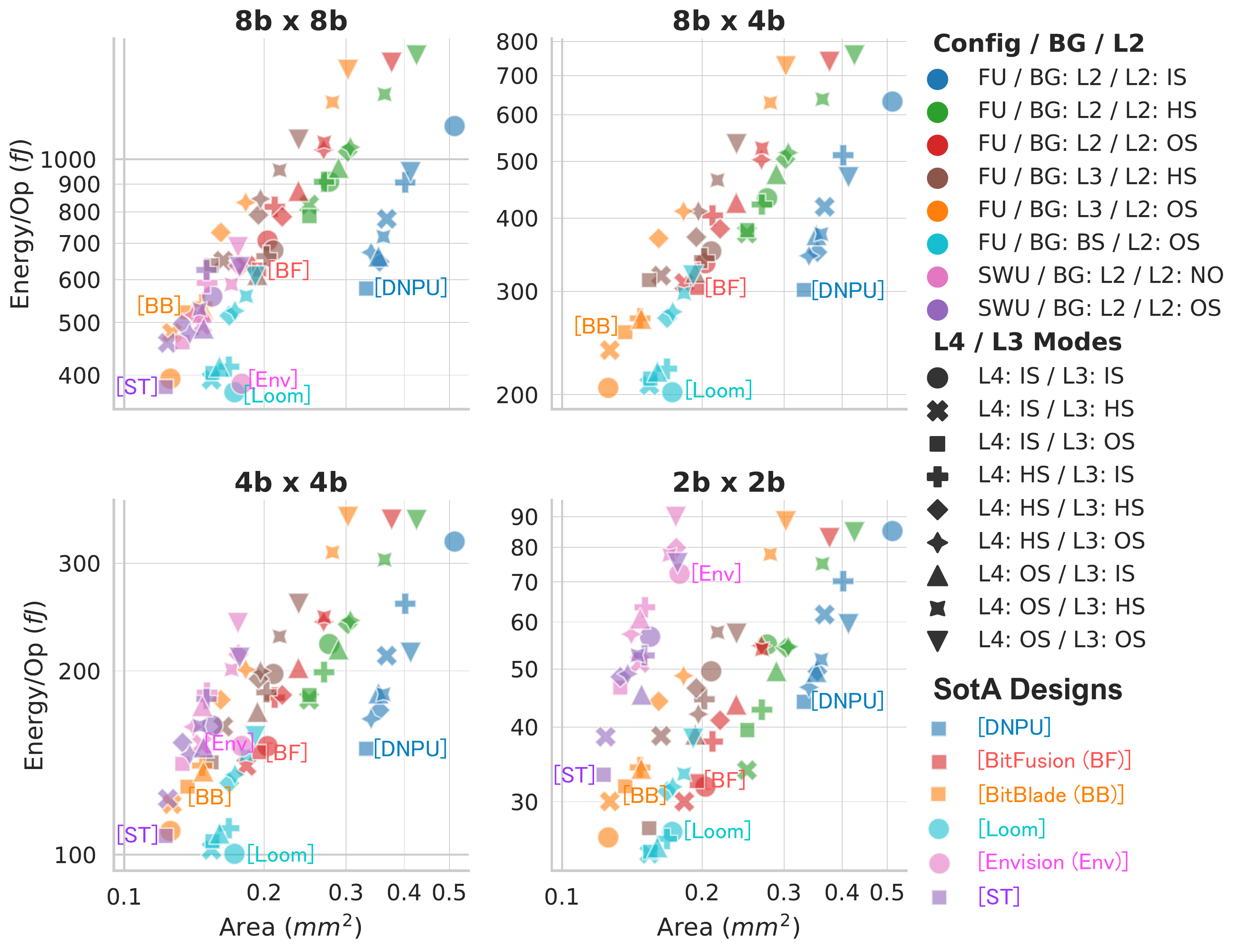}
      \label{subfig:Scatter-1}
    }
    \caption{Energy/Op ($fJ$) vs. Area ($mm^2$) at (a) 200 MHz, and (b) 1 GHz. Both X and Y axis are log-scale.}
    \label{fig:Scatter-all}
  \end{figure*}

  \subsection{Energy Efficiency}
  All supported designs are compared in terms of energy per operation, with one operation defined as one full multiplication or one addition (i.e., one full MAC operation is defined as two operations). The resulting energy efficiency heatmaps are shown for 200 MHz at \Cref{subfig:Heatmap-0.2} and 1 GHz at \Cref{subfig:Heatmap-1}. Additionally, the designs are compared with symmetric (8b$\times$8b, 4b$\times$4b, 2b$\times$2b) and one asymmetric (8b$\times$4b) precisions. Note that SWU designs do not support asymmetric precision scalability. Each row denote an L4/L3 mode, while each column denote a configuration/BG/L2 combination. 
  
  At 200 MHz (\Cref{subfig:Heatmap-0.2}), the most energy efficient columns across different precisions are the (FU / BG: L3 / L2: OS), (FU / BG: BS / L2: OS), and (SWU / BG: L2 / L2: OS) configurations. At full precision, the (SWU / BG: L2 / L2: OS) designs are more efficient. This can be attributed to the simpler design and non-configurable shifters, while still maintaining the same throughput as FU. At lower precisions, SWU have lower number of operations, and thus become more in line with FU designs. Intuitively, one can understand that (FU / BG: L3) and (FU / BG: BS) designs should be more efficient than (FU / BG: L2) designs, due to the fact that the shifter's overhead is shared across the L2 units in this case. BS designs have an additional switching overhead of the internal registers, and since here the frequency is relatively low, there is not much gain in return.
  
  The key factor for good energy efficiency is to make L2 OS, as it produces only one partial product regardless of precision, thus simplifying L3 and L4 designs. At 200 MHz, L3 and L4 unrollings don't have as much impact as long as L2 is OS. The best L4/L3 configurations in that case are IS/IS, IS/OS, OS/IS. If both L4 and L3 swing towards the OS side, the critical path gets longer due to the adder trees, and as a result the energy efficiency is negatively impacted, though not by much since this is still at a low clock frequency.
  
  Fig. \ref{subfig:Heatmap-1} shows a similar trend at 1 GHz clock frequency, with some small differences. (L2: OS) is still a key factor for good energy efficiency, and (BG: L3) and (BG: BS) are still more efficient then (BG: L2). However, (BG: BS) has a slight edge over (BG: L3) due to the reduced critical path, as BS designs' internal registers start to show their benefit at higher clock frequencies. In line with the 200 MHz results, the best L4/L3 configurations are IS/IS, IS/OS, OS/IS. Having all level unrollings as OS yields a detrimental effect on the energy efficiency, as the critical path gets longer, conflicting with the tight timing requirement.
  
  \begin{figure*}[p]
    \centering
    \subfloat{
      \includegraphics[width=0.9\textwidth]{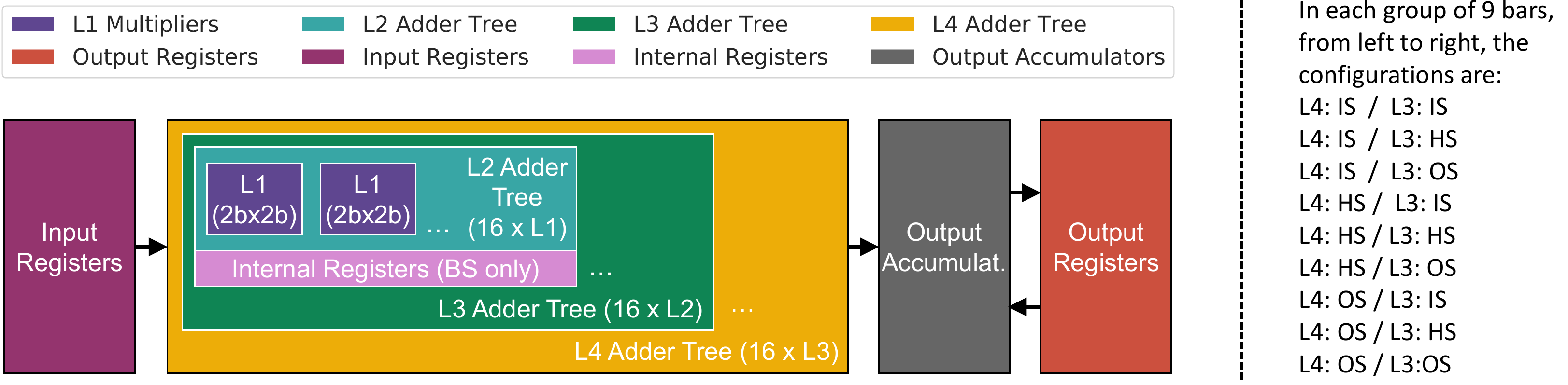}
    } 
    \caption{Legend illustration for \Cref{fig:breakdown_200MHz} and \Cref{fig:breakdown_1GHz}. Input Registers contain input activations and weights.}
    \label{fig:legend}
    \subfloat[Energy/Op ($fJ$) for 8b$\times$8b]{
      \includegraphics[width=0.48\textwidth]{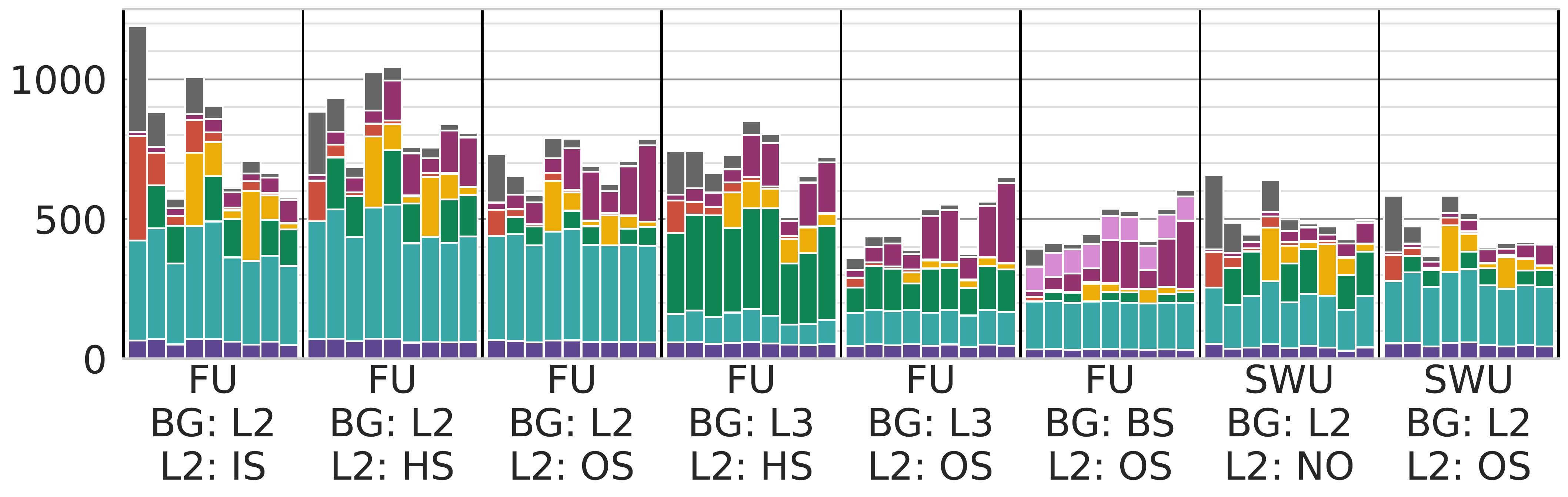}
    }
    \subfloat[Energy/Op ($fJ$) for 8b$\times$4b]{
      \includegraphics[width=0.48\textwidth]{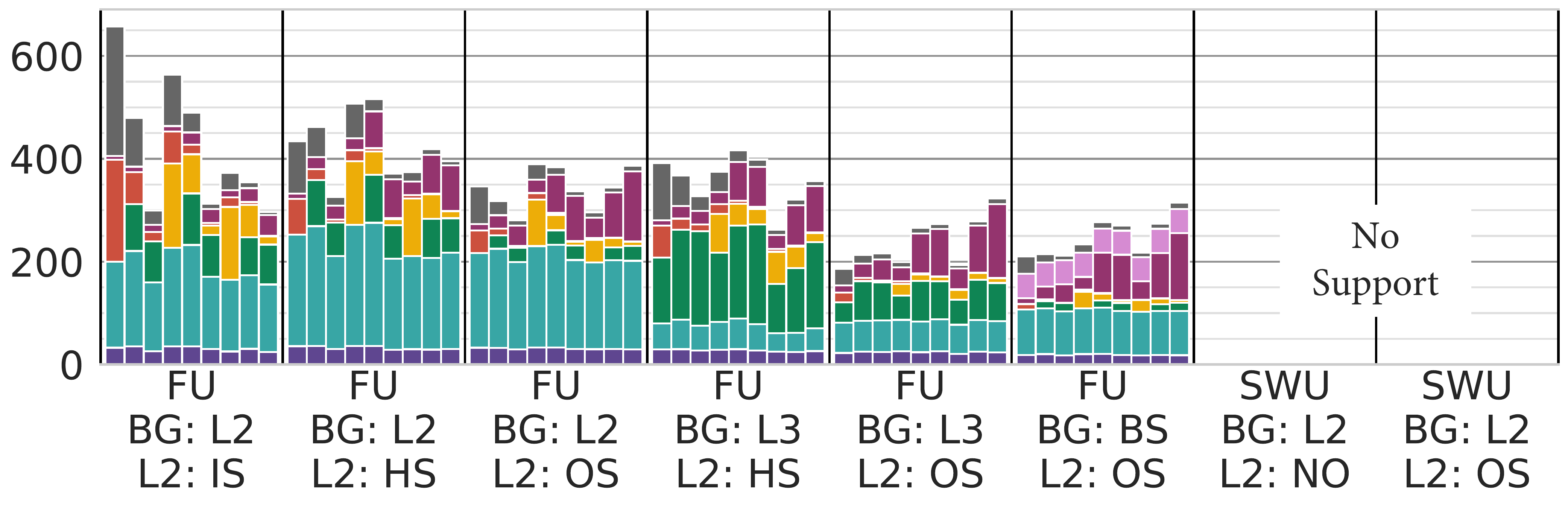}
    } \par
    \subfloat[Energy/Op ($fJ$) for 4b$\times$4b]{
      \includegraphics[width=0.48\textwidth]{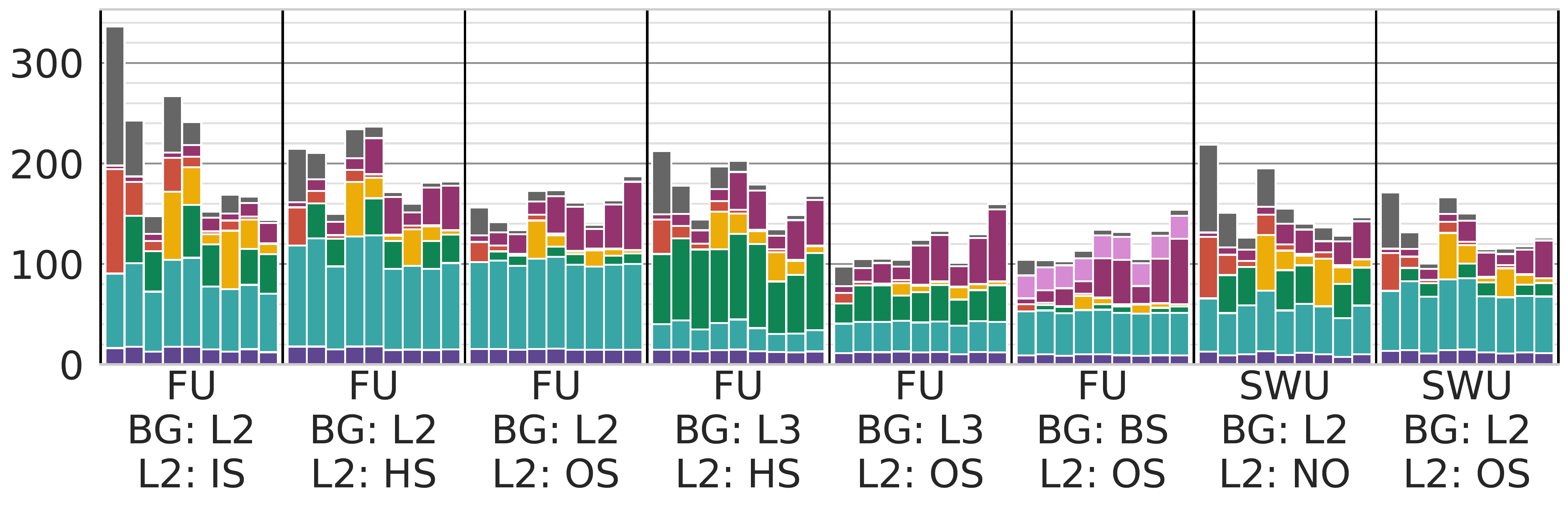}
    } 
    \subfloat[Energy/Op ($fJ$) for 2b$\times$2b]{
      \includegraphics[width=0.48\textwidth]{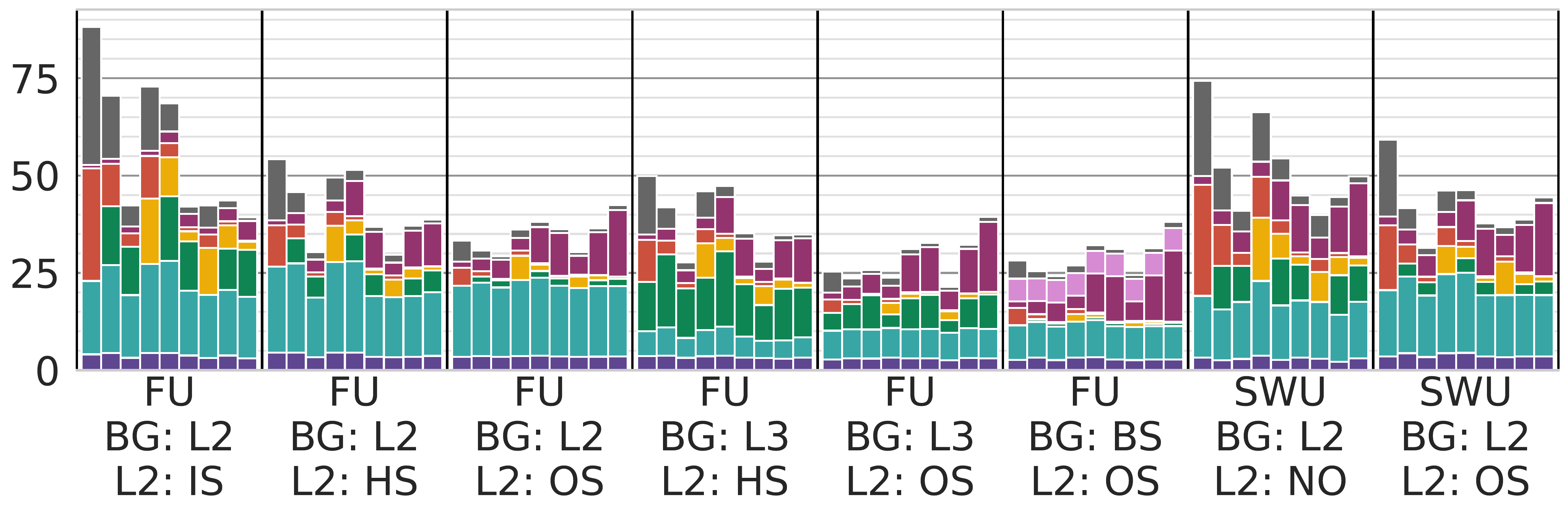}
    }
    \caption{Energy/Operation for 200 MHz.}
    \label{fig:breakdown_200MHz}
    
    \subfloat[Energy/Op ($fJ$) for 8b$\times$8b]{
      \includegraphics[width=0.48\textwidth]{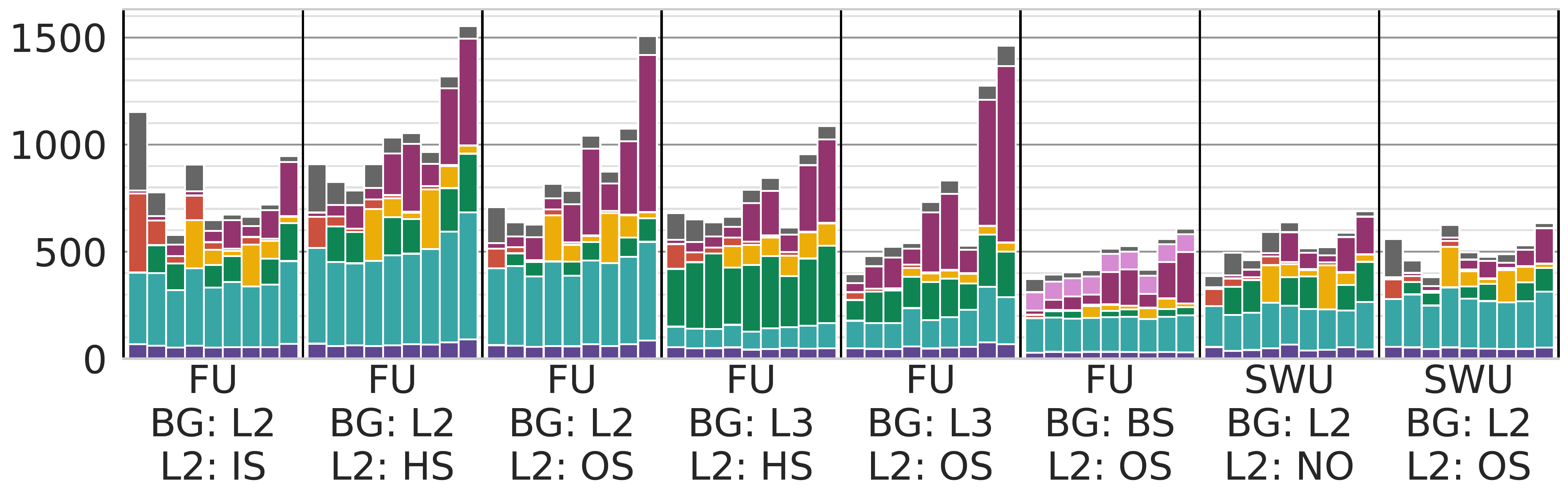}
    }
    \subfloat[Energy/Op ($fJ$) for 8b$\times$4b]{
      \includegraphics[width=0.48\textwidth]{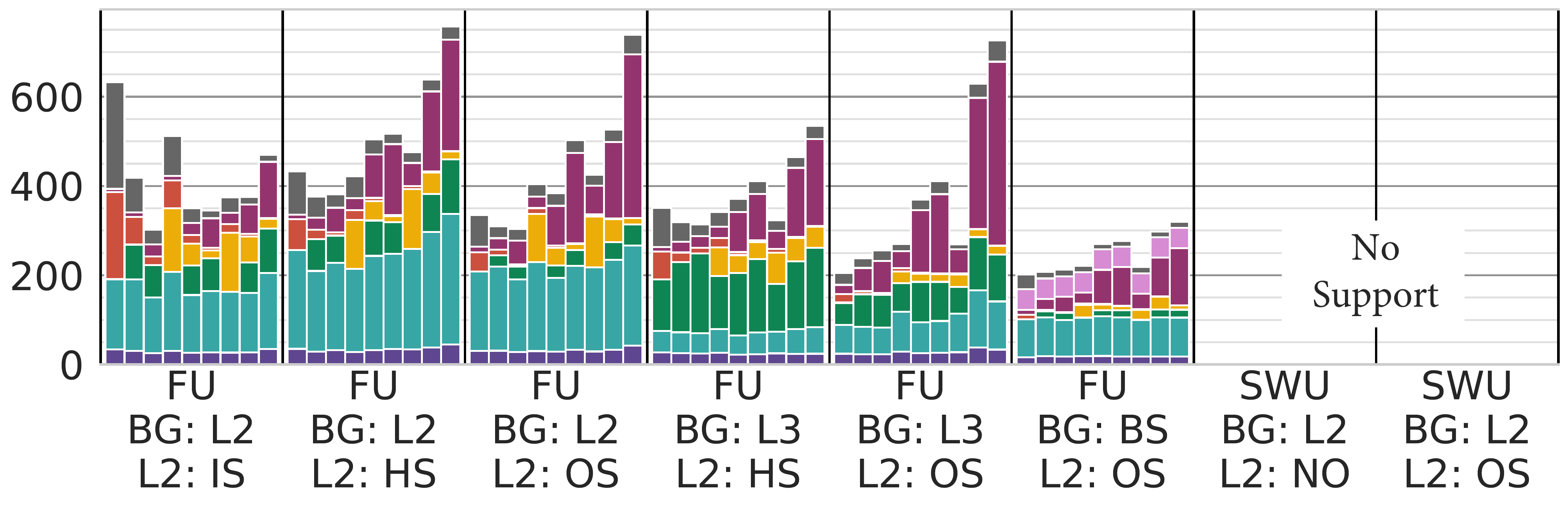}
    } \par
    \subfloat[Energy/Op ($fJ$) for 4b$\times$4b]{
      \includegraphics[width=0.48\textwidth]{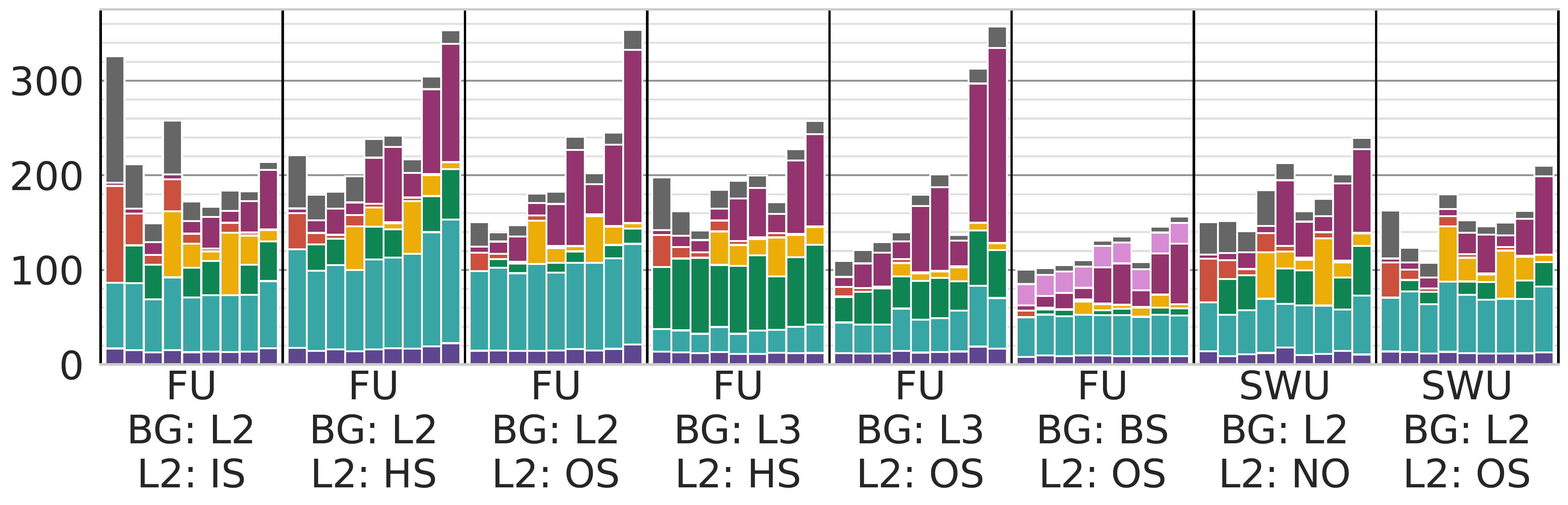}
    } 
    \subfloat[Energy/Op ($fJ$) for 2b$\times$2b]{
      \includegraphics[width=0.48\textwidth]{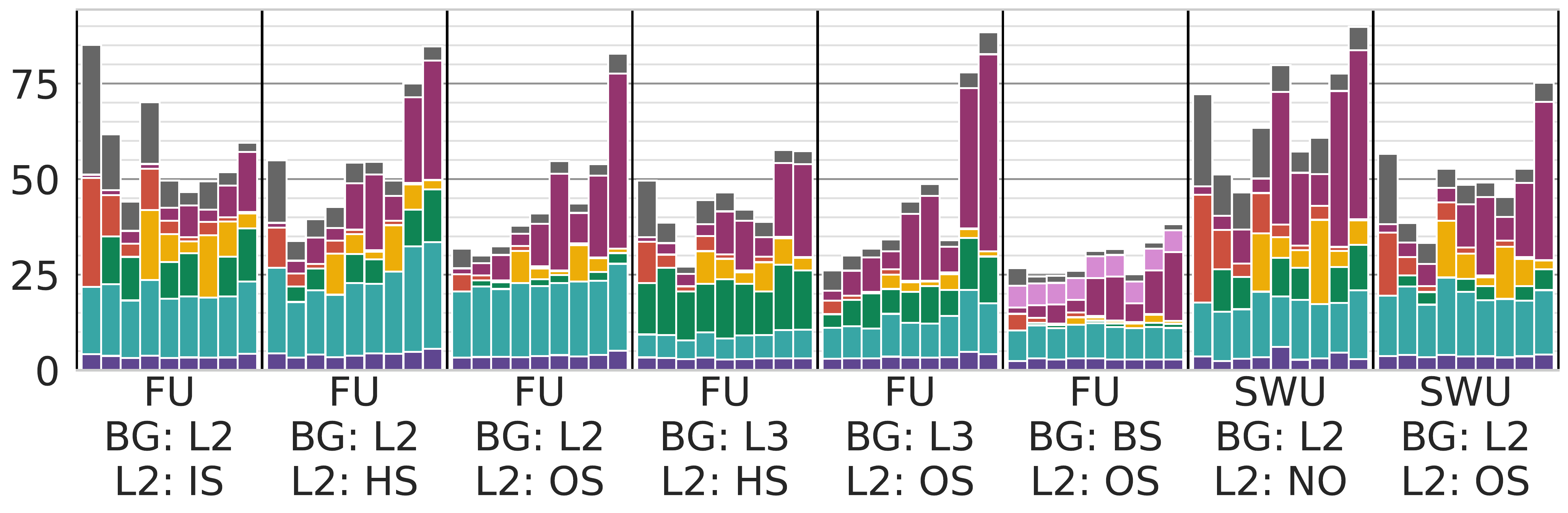}
    } 
    \caption{Energy/Operation for 1 GHz.}
    \label{fig:breakdown_1GHz}
    
    \subfloat[Area ($mm^2$) for 200 MHz]{
      \includegraphics[width=0.48\textwidth]{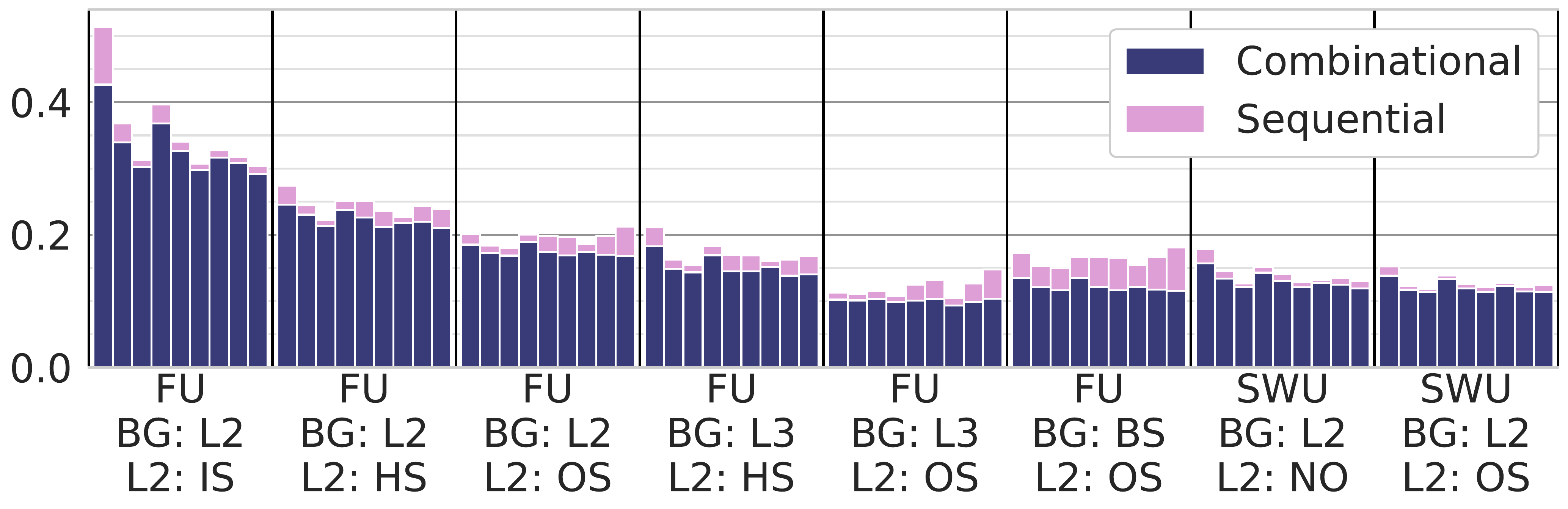}
      \label{subfig:area_200MHz}
    }
    \subfloat[Area ($mm^2$) for 1 GHz]{
      \includegraphics[width=0.48\textwidth]{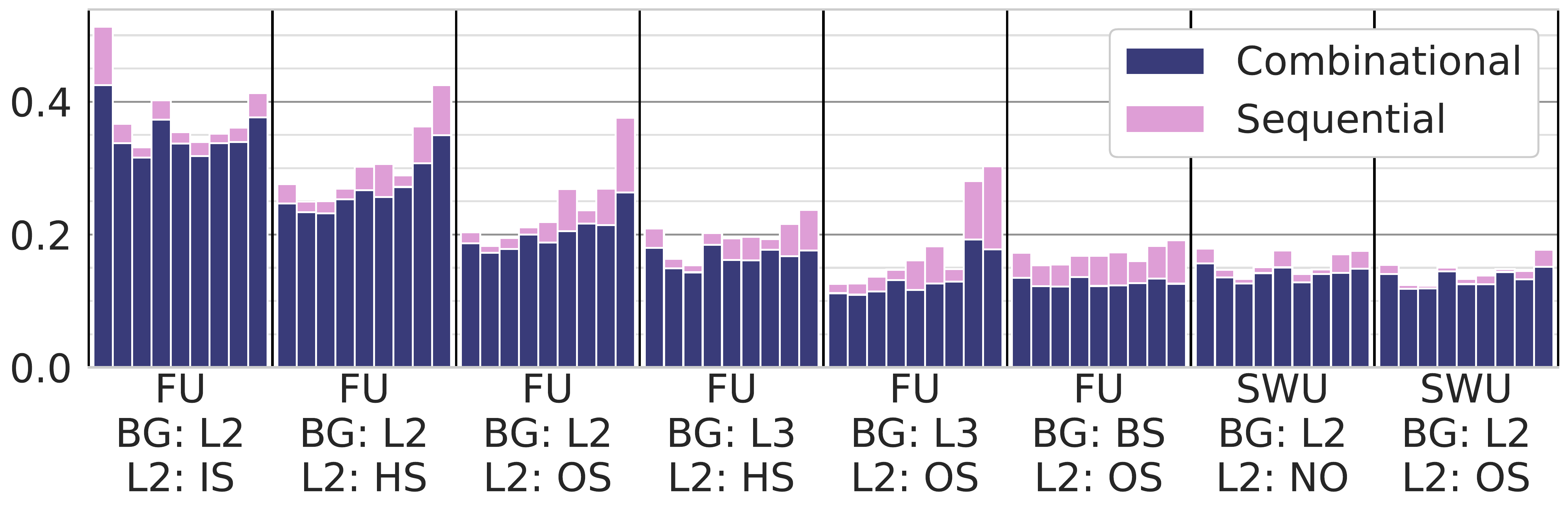}
      \label{subfig:area_1GHz}
    }
    \caption{Area breakdown.}
    \label{fig:Area}
  
  \end{figure*}
  
  The stellar performance of BS designs in this benchmark are in stark contrast to \cite{camus2019review}, which showed BS MAC units as the worst performer.
  The primary reason for the improved energy efficiency is that, in this work, we assume “L2 is OS” for all BS designs in the PSMA template. As such, the hardware overhead of the internal registers is amortized across the L1 units within each L2 unit (i.e., BS-L2), which ultimately reduces the energy/op. Contrarily, in \cite{camus2019review}, the study is done at single MAC unit level, and thus the amortization opportunity is ignored (i.e., BS-L1). This result highlights the importance of hardware resource sharing in array-level BS precision-scalable MAC engine designs.
  
  \subsection{Energy vs. Area}
  To get a better assessment of the most efficient array, the benchmarked designs are compared in terms of energy per operation vs. area. Fig. \ref{subfig:Scatter-0.2} and \Cref{subfig:Scatter-1} show scatter plots of energy/operation vs. area for 200 MHz and 1 GHz, at different symmetric and asymmetric precisions. Each color represents a different (config / BG / L2) combination, while the each shape represent a different (L4 / L3) modes. The most optimal designs lie in the bottom-left corner.
  
  At 200 MHz, the optimal designs are the purple and orange ones, which represent (SWU / L2: OS) and (FU / BG: L3 / L2: OS).  At 1 GHz, the cyan colored markers (BS) have better energy efficiency, while the purple and orange ones have smaller areas. As will be shown later, this can be attributed to the lack of internal registers in the (SWU / L2: OS) and (FU / BG: L3 / L2: OS) designs compared to BS designs. In most cases, the blue markers (FU / BG: L2 / L2: IS) are on the upper right corner of the graph, which makes them one of the least efficient configurations. This can be attributed to the fact that having IS at the lowest level of the array produces lots of independent partial products at lower precisions, which cannot be added together. Generally, it can be seen that the top left and bottom right corners are mostly empty, which indicates that in most cases, there's a strong correlation between area and energy per operation in this benchmark.

  \subsection{Breakdown}
  To gain more insights about why some array configurations perform better than the others, a breakdown of the energy per operation and area for all PSMA configurations are summarized in \Cref{fig:breakdown_200MHz,fig:breakdown_1GHz,fig:Area} for 200 MHz and 1 GHz with \Cref{fig:breakdown_200MHz,fig:breakdown_1GHz}'s legend shown in \Cref{fig:legend}. The breakdowns include the energy contribution of L1 multipliers, adder trees (in L2, L3, and L4), final accumulation adders, and registers (input, output, and BS internal). 
  
  The L1 2-bit multipliers energy consumption is nearly identical in all cases, and only contributes a small fraction of the total energy consumption. L2 adder trees consume more energy if BG is unrolled at L2, since in this case the adder tree include both adders and shifters. L2 adder tree's energy for SWU designs is a bit lower than for FU designs due to the non-configurable shifters. The complexity of the L2 adder trees goes significantly down as we move to BG in L3 or BS designs. If L3 or L4 are IS, it means that none of the partial results from the previous level get added together, thus there would be no L3/L4 adder trees in this case. Additionally, L3 adder tree (if it exists) consume more energy whenever L2 is IS as the precision gets lower.
  
 Output registers and accumulation adders dominate energy consumption as more levels become IS, while input register's energy is much lower, and vice versa. Internal registers only exist in BS designs, and it consumes a sizeable amount of the total energy consumption of BS designs. At 1 GHz, the arrays where most levels are OS yield higher energy consumption than other designs. The reason for that is that OS arrays have long critical paths, since the partial results have to go through complex adder trees in most of the levels. To meet the timing constraint, the energy consumption gets negatively impacted. 
  
  Area breakdowns are shown in \Cref{fig:Area}. As shown in the figures, the area is dominated by the combinational logic, with the sequential logic only having a small contribution to the total area. Additionally, all designs where L2 is OS have lower area than their ``L2: IS" and ``L2: HS" counterparts. That's because as L2 generate more partial products, L3 and L4 adder trees become more complicated to ensure L2 partial results are not added together. In general, a larger area is observed for 1 GHz frequency, and this can be attributed to larger cell sizes in order to meet the timing constraints of the architectures, except for the BS designs.

\section{Conclusions}\label{sec:Conclusion}
  This work proposes a new loop representation extending the traditional 7 CNN for-loops with additional Bit-Group loops for precision-scalability. Additionally, the traditional 7 CNN for-loops are categorized into three different categories (IS, WS, and OS loops). This allows the introduction of a new taxonomy which exploits the flexibility offered by the newly proposed CNN loop categories, where the BGs can be unrolled spatially at different hierarchical levels of the MAC array or temporally (bit-serial). Afterwards, existing SotA PSMAs are mapped to the introduced taxonomy. The SotA mapping is accompanied by a deep discussion on the effects of each taxonomy parameter on the underlying hardware, and how they affect the input and output bandwidths of the PSMAs. A new uniform and highly parameterized PSMA template is introduced that covers a large subset of the design space spanned by the new taxonomy, and supports symmetrical and asymmetrical precision scaling of inputs and weights.
  
  All the architectures in the constrained design space are synthesized using TSMC 28 nm technology, and benchmarked extensively in terms of energy efficiency and area. From the conducted benchmarks, it is shown that BG unrolling in L2 is the least ideal case (e.g. Bitfusion \cite{sharma2018bitfusion}, DNPU \cite{shin2017dnpu}), and it's better to have BG unrolling at L3 for lower frequencies (e.g. BitBlade \cite{ryu2019bitblade}, Ghodrati \cite{ghodrati2020bit}) or temporally unrolling BGs for higher frequencies (e.g. Loom \cite{sharify2018loom}, Stripes \cite{judd2016stripes}). The benefit of having BG unrolled in L3 or temporally is that the shifters are amortized across different L2 units, making the adder trees less complex. It's generally a good idea to have a mixture of IS/WS and OS loops throughout the array levels. FU designs are better suited for workloads where lower precisions are common, whereas SWU designs are better suited for higher precisions dominated workloads. 
  
  According to the benchmark, (L2: OS) is the key factor for energy efficient PSMAs. At 200~MHz, (BG: L3) is slightly better than (BG: BS) as the internal registers of BS impose an overhead with no extra benefit at lower frequencies. At 1~GHz, BS designs have a slight edge over (BG: L3) in terms of energy efficiency, while (BG: L3) are better in terms of area. The best L4/L3 unrolling in both cases are IS/IS (Loom-like \cite{sharify2018loom}), IS/OS (BitBlade-like \cite{ryu2019bitblade}), and OS/IS. The best performing SWU design across 200 MHz and 1 GHz is (L4: IS, L3: OS, L2: OS), which is an ST-like architecture \cite{mei2019subword}.
  
  One thing to be noted is that, in this work's benchmarks, the best PSMAs were chosen based on an ideal workload, which guarantees full hardware utilization for all designs across all precisions. However, different workloads may affect the utilization of the benchmarked designs, leading to a different performance landscape. Assessing the effects of different non-ideal workloads on different PSMAs is left for future works. 
  
  All in all, deep learning models have proven to be resilient to lower precisions, which in turn propelled the research in PSMA design for DNN accelerators. This work enabled to better categorize current SotA PSMAs, to clearly understand the trade-offs within/between each design, and to insightfully find the optimal PSMA under different circumstances.

\section*{Acknowledgment}
The authors would like to thank Vincent Camus for his insights on the automation of the benchmarking process.

\ifCLASSOPTIONcaptionsoff
  \newpage
\fi


\begin{IEEEbiography}[{\includegraphics[width=1in]{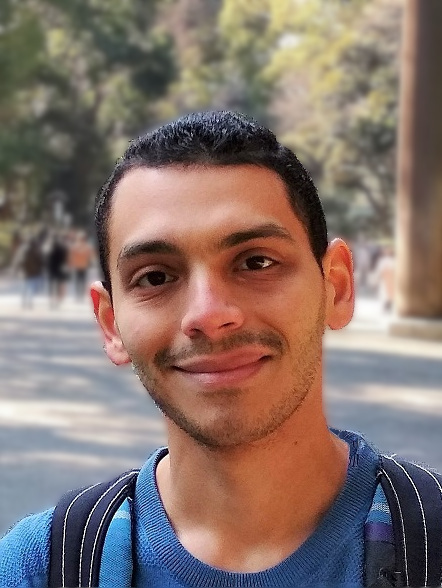}}]{Ehab M. Ibrahim}
  received the B.Sc. degree in Electronics and Communications Engineering from Zagazig University, Zagazig, Egypt, in 2012, and the M.Sc. degree in Electronics and Communications Engineering from Egypt-Japan University of Science and Technology (E-JUST), Egypt, in 2019. He was an intern at the National Institute of Informatics, Tokyo, Japan in the spring of 2019. He's currently a Digital Integrated-Circuits designer at Magics Technologies, Belgium. This work was done when he was part of MICAS, KU Leuven as a research assistant from 2020 to 2021. His current interests include energy-efficient deep learning acceleration, hardware-software co-design, and precision-scalable computing.
\end{IEEEbiography}
\vskip -2\baselineskip plus -1fil
\begin{IEEEbiography}[{\includegraphics[width=1in]{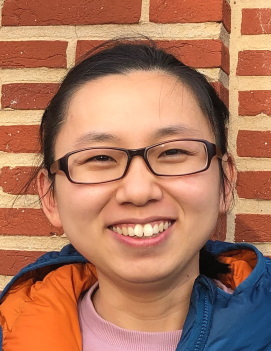}}] {Linyan Mei} received B.Sc. in Electronic Science and Technology from Beijing Institute of Technology (BIT), China, in 2016, and M.Sc. in Electrical Engineering from KU Leuven, Belgium, in 2018. She is currently pursuing a Ph.D. on digital design for embedded machine-learning processors with the MICAS laboratories, KU Leuven.
She was an Intern with IMEC, Leuven, Belgium, from 2017 to 2018; an Intern with Facebook, Menlo Park, U.S., in the winter of 2021. Her current research interests include design space exploration for energy-efficient deep neural network accelerators, algorithm-hardware-mapping co-optimization, and precision-scalable computing.\end{IEEEbiography}
\vskip -2\baselineskip plus -1fil
\begin{IEEEbiography}[{\includegraphics[width=1in]{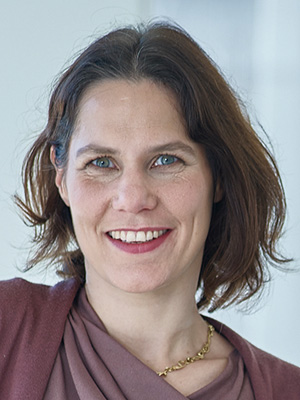}}]{Marian Verhelst}
  received the Ph.D. degree from KU Leuven, Belgium, in 2008. She was a Visiting Scholar with the Berkeley Wireless Research Center of the University of California Berkeley in the summer of 2005, and was a Research Scientist with Intel Labs, Hillsboro, OR, USA, from 2008 until 2011. She is currently an Full Professor with the MICAS  Laboratories, Electrical Engineering Department, KU Leuven. Her research focuses on embedded machine learning, hardware accelerators, self-adaptive circuits and systems, sensor fusion, and low-power edge processing. She is a member of the DATE and ISSCC executive committees, is the TPC Co-Chair of AICAS 2020 and tinyML 2020, and a TPC member of DATE and ESSCIRC. She is an SSCS Distinguished Lecturer, was a member of the Young Academy of Belgium, a member of the STEM advisory committee to the Flemish Government, and an Associate Editor for TVLSI, TCAS-II, and JSSC. She currently holds a prestigious ERC Starting Grant from the European Union and was the laureate of the Royal Academy of Belgium in 2016.
\end{IEEEbiography}

\end{document}